\documentclass[11pt]{article} 
\usepackage[top=1.25in, bottom=1.25in, left=1in, right=1in]{geometry}
\usepackage[sectionbib]{natbib}
\usepackage{array,epsfig,fancyheadings,rotating}
\usepackage{fancyhdr}
\usepackage{titling}
\usepackage[]{hyperref}  
\usepackage{sectsty, secdot}
\sectionfont{\fontsize{11}{14pt plus.8pt minus .6pt}\selectfont}
\renewcommand{\theequation}{\thesection\arabic{equation}}
\subsectionfont{\fontsize{11}{14pt plus.8pt minus .6pt}\selectfont}

\usepackage{amsmath}
\usepackage{amssymb}
\usepackage{amsfonts}
\usepackage{multirow}
\usepackage{amsthm}
\usepackage{algorithm}
\usepackage{algpseudocode}
\usepackage{amsmath}
\usepackage{natbib}
\usepackage{hyperref}
\hypersetup{colorlinks,citecolor=blue,urlcolor=blue,filecolor=blue,backref=page}
\usepackage{graphicx}
\usepackage{mathtools}
\usepackage{float}
\usepackage{setspace}

\theoremstyle{plain}
\newtheorem{thm}{Theorem}[section]
\usepackage{color}

\author{Canida T. and Ke H. and Ma T.}

\begin{document}


\markright{ \hbox{\footnotesize\rm Statistica Sinica
}\hfill\\[-13pt]
\hbox{\footnotesize\rm
}\hfill }

\markboth{\hfill{\footnotesize\rm FIRSTNAME1 LASTNAME1 AND FIRSTNAME2 LASTNAME2} \hfill}
{\hfill {\footnotesize\rm Travis Canida and Hongjie Ke and Tianzhou Ma} \hfill}

\renewcommand{\thefootnote}{}
$\ $\par


\fontsize{11}{11pt plus.8pt minus .6pt}\selectfont \vspace{0.8pc}
\begin{center}
\bfseries\large
Multivariate Bayesian variable selection with application to multi-trait genetic
fine mapping
\end{center}
\vspace{.4cm} \centerline{Travis Canida, Hongjie Ke, Shuo Chen, Zhenyao Ye and Tianzhou Ma} \vspace{.4cm} \centerline{\it
 University of Maryland} \vspace{.55cm} \fontsize{9}{11.5pt plus.8pt minus
.6pt}\selectfont


\begin{quotation}
\noindent {\it Abstract:}

Variable selection has played a critical role in modern statistical learning and scientific discoveries. Numerous regularization and Bayesian variable selection methods have been developed in the past two decades for variable selection, but most of these methods consider selecting variables for only one response. As more data is being collected nowadays, it is common to analyze multiple related responses from the same study. Existing multivariate variable selection methods select variables for all responses without considering the possible heterogeneity
across different responses, i.e. some features may only predict a subset of responses but not the rest. Motivated by the multi-trait fine mapping problem in genetics to identify the causal variants for multiple related traits, we developed a novel multivariate Bayesian variable selection method to select critical predictors from a large number of grouped predictors that target at multiple correlated and possibly heterogeneous responses. Our new method is featured by its selection at multiple levels, its incorporation of prior biological knowledge to guide selection and identification of best subset of responses predictors target at. We showed the advantage of our method via extensive simulations and a real fine mapping example to identify causal variants associated with different subsets of addictive behaviors.

\vspace{9pt}
\par
\end{quotation}\par

\def\thefigure{\arabic{figure}}
\def\thetable{\arabic{table}}

\renewcommand{\theequation}{\thesection.\arabic{equation}}

\fontsize{11}{11pt plus.8pt minus .6pt}\selectfont

\setcounter{section}{1} 
\setcounter{equation}{0} 
\noindent {\bf 1. Introduction}
\label{sec:intro}




\par

Variable selection has been a compelling problem in statistical modeling and played a central role in modern statistical learning and scientific discoveries in diverse fields. With the advancement of technology in recent years, high-dimensional data with a huge number of features become rules rather than exception. Classical best subset selection methods such as those using AIC and BIC are computationally too expensive for most modern statistical applications. Penalized regression or regularization methods have been common choices for modern variable selection by including different forms of penalty functions in regression models to shrink regression coefficients towards zeros. For example, lasso \citep{tibshirani1996regression} penalizes on the L1-norm of the regression coefficients while elastic net \citep{zou2005regularization} induces a linear combination of L1 and L2 penalties. As group structure arises naturally in many applications, the group lasso generalizes lasso to select grouped variables for accurate prediction in regression \citep{yuan2006model} and \cite{simon2013sparse} further extended to sparse group lasso for within group selection. On the Bayesian side, inspired by the hierarchical structure of Laplace prior, \cite{park2008bayesian} proposed a Bayesian formulation of lasso and \cite{casella2010penalized} further derived the group version Bayesian lasso and elastic net. Alternatively, people introduced a sparsity-induced ``spike-and-slab'' prior, consisting of a point mass at zero or centered around zero with small variance (``spike'') and a diffuse uniform or large variance distribution (``slab''), on the regression coefficients to achieve variable selection \citep{mitchell1988bayesian,george1993variable,kuo1998variable,ishwaran2005spike}. One may refer to \cite{o2009review} for a complete review of spike-and-slab prior based methods and their variants. \cite{xu2015bayesian,hernandez2013generalized,zhang2014bayesian,chen2016bayesian,zhu2019bayesian} have later extended the spike-and-slab priors for variable selection at group level and within groups. The aforementioned methods mainly deal with a single response, as more data is being collected nowadays, multiple related responses become available. For example, studies typically analyze multiple clinically relevant outcomes such as blood pressure, cholesterol and low-density lipo-protein levels together to assess one's overall risk for cardiovascular disease. Running separate regression for each response ignores their correlation so a multivariate regression model is often recommended. Both penalized regression and Bayesian variable selection methods have been developed to select variables in multivariate regression model \citep{peng2010regularized,li2015multivariate,liquet2017bayesian}. However, the above methods select variables related to all responses without considering the possible heterogeneity across responses, i.e. some features may only predict a subset of responses but not the rest.

\begin{figure}[t] 
\centering
\includegraphics[scale=0.6]{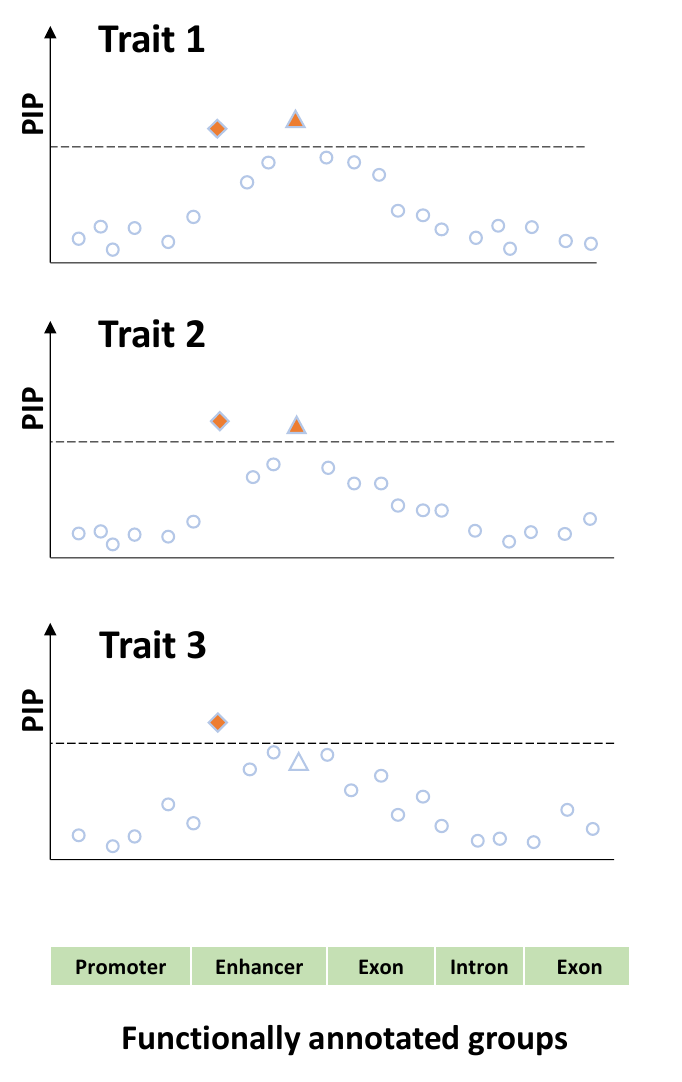}
\caption{A multi-trait fine mapping example that motivates the development of our method. Y-axis refers to the posterior inclusion probability (PIP) and X-axis refers to the position of SNPs on the genome. The two causal variants are symbolized by diamond or triangle. The dashed line indicates a PIP threshold above which a variant can be regarded as causal to that trait. Variants are shaded orange when they pass the PIP threshold.}
\label{fig:1}
\end{figure}

Our method, though potentially of broad interest, has been mainly motivated by the multi-trait fine mapping studies in genetics \citep{kichaev2017improved}. Genetic fine mapping aims to pinpoint the causal variants (i.e. single nucleotide polymorphisms or SNPs) associated with the trait in local genomic regions determined by genome-wide association studies (GWAS) \citep{schaid2018genome}. Fine mapping can often be framed as a variable selection problem that aims to identify a parsimonious set of variants from a large number of correlated SNPs possibly with group structure. Figure \ref{fig:1} shows the Manhattan plots of posterior inclusion probability (PIP; a higher PIP indicates a higher probability of the variant being causal in fine mapping) of SNPs in a local genomic region potentially associated with three correlated traits. The SNPs in this region can be divided into five groups depending on their functional annotation (e.g. promoter, enhancer, exon and intron) or linkage disequilibrium (LD) patterns. There are two causal variants located in the 2nd group: the first variant in diamond is causal to all three traits (above PIP threshold indicated by the dashed line), while the second variant in triangle is a causal to trait 1 and 2 but not 3. Such phenomenon of a variant targeting at multiple traits (also known as ``pleiotropy'') is common in genetic studies \citep{paaby2013many}, and it is also common to see that some variants are causal to only a subset of traits but not all traits under study. For example, recent imaging genetic studies have identified a few common genetic risk variants of cognitive traits and some imaging phenotypes but not other imaging phenotypes \citep{zhao2019heritability,zhao2021common}. 

In this paper, we developed a novel multivariate Bayesian variable selection method well suited to settings when there are a large number of grouped predictors but only a few with detectable effects on multiple correlated responses. These are typical characteristics of multi-trait genetic fine mapping, where we aim to identify a few causal variants, from a large number of highly correlated and grouped SNPs, for multiple closely related phenotypes. Our method is built upon existing Bayesian variable selection methods first proposed by \cite{kuo1998variable} as an alternative reparameterization of spike-and-slab prior, and later developed by \cite{sillanpää2005bayesian,sillanpää2006association,xu2015bayesian,zhu2019bayesian} for group-level selection with wide genetic and multi-omics applications. \cite{liquet2017bayesian} proposed multivariate sparse group selection method with spike-and-slab prior to select predictors with group structure associated with several correlated outcome variables. \cite{wang2020simple} proposed a Bayesian analogue of traditional stepwise selection methods with potential extension to select variables simultaneously for multiple outcomes. Other Bayesian methods more specifically designed for fine mapping were also developed to identify pleiotropic SNPs that are potentially causal variants of multiple traits \citep{stephens2013unified,hormozdiari2016colocalization,kichaev2017improved,schaid2018genome}. Several important issues when applied to multi-trait fine mapping, however, are not well addressed in these methods. First, all these methods assume common predictors of all the responses. In multi-trait fine mapping, heterogeneity usually exists so a variant may be causal to some traits but not the others. Secondly, as the dimension of response grows, it becomes an imperative task to identify subset(s) of responses that are most impacted by the predictor which might help generate new hypotheses. For example, in fine mapping causal variants of neuroimaging phenotypes, it is appealing to identify the subset of neuroimaging phenotypes (possibly in spatial proximity and with similar functional roles) causal SNPs target at to improve our understanding of the heterogeneous genetic effect on the brain. Lastly, incorporation of background knowledge to guide variable selection has been repeatedly emphasized in genomic application and is especially critical in fine mapping. For example, the group structure as defined by physical location, functional annotation, haplotype or LD patterns is useful in narrowing down the selection into smaller sets of functionally relevant SNPs; the probability of a variant being causal is related to existing knowledge stored in reference database about the variant, e.g. whether the variant is an expression quantitative trait loci (eQTL) or other multi-omics QTL \citep{schaid2018genome}. Incorporating multiple lines of evidence of prior knowledge will help improve causal variant selection performance and draw more reliable interpretation and conclusion.

Comparing to existing methods, our method has several advantages that provide additional insights into the multi-trait fine mapping problem: (1) our method conducts variable selection at both individual feature and group levels, and the selection is specific to each response, thus allowing the selection of causal variants of multiple heterogeneous traits from function or haplotype defined groups of SNPs in fine mapping; (2) we introduce a new concept of subset PIP in multi-trait fine mapping to select subset(s) of traits a causal variant targets at; (3) under the full Bayesian framework, our method is flexible in incorporating prior biological knowledge into the fine mapping problem to guide grouping and weight the probability of causal variants. In addition, we showed the posterior median estimator of the regression coefficients in our model is a soft thresholding estimator with selection consistency and asymptotic normality. We performed extensive simulations comparing our proposed method to existing multivariate regularization,  Bayesian variable selection and fine-mapping methods. We applied our method to a multi-trait fine mapping problem in UK Biobank (UKB) and identified several important variants that target at different subsets of addictive behaviors and their risk factors. The paper is organized as follows. In Sections \ref{sec:review}, \ref{sec:method} and \ref{sec:related_methods}, we review the development of Bayesian variable selection methods, propose our model and its key features, investigate its theoretical properties, propose a new concept of subset PIP for posterior inference of our method, and compare to existing methods. In Section \ref{sec:simulation}, we show our model's performance in extensive simulations. In Section \ref{sec:application} we show a real data example and in Section \ref{sec:discussion} we provide discussion on the potential extension of our method in future studies.

\section{Review of Bayesian variable selection methods}
\label{sec:review}

We first consider a linear regression model for a univariate response $Y_i$ with $p$ covariates $X_{ij}, j=1,\ldots,p$: 
\begin{equation}
Y_i=\sum_{j=1}^{p}\beta_{j}X_{ij}+\epsilon_i, \quad \epsilon_i \sim N(0,\sigma^2),
\end{equation}
for samples $i=1,2,\ldots,n$. The data are assumed to be centered so the intercept can be ignored. The spike and slab model \citep{mitchell1988bayesian,george1993variable}, among the most popular Bayesian variable selection methods, imposes the following prior on each regression coefficient $\beta_j$:  
\begin{equation}
\begin{split}
\beta_j &  \sim (1-\gamma_j) N(0,\tau_j^2) +\gamma_j N(0,c_j^2\tau_j^2); \\ 
\gamma_j & \sim Bern(\pi_j),
\end{split}
\end{equation}
where $\tau_j^2$ is usually chosen very small, and $c_j^2$ is usually chosen large. $\gamma_j$ indicates whether the $j$th covariate is predictive of the outcome: when $\gamma_j=1$,  $\beta_j$ is drawn from the wide slab part $N(0,c_j^2\tau_j^2)$; when $\gamma_j=0$, $\beta_j$ is drawn from the close-to-zero spike part. This prior will encourage a sparse model shrinking coefficients under spike towards zero.  

\cite{kuo1998variable} proposed an indicator variable selection model as an alternative to spike and slab model. The idea is to attach an indicator variable to each coefficient for model selection:  
\begin{equation}
\begin{split}
Y_i & =\sum_{j=1}^{p}\beta_j\gamma_j X_{ij}+\epsilon_i,\epsilon_i\sim N(0,\sigma^2); \\ 
\beta_j & \sim N(0,s^2), \qquad \gamma_j \sim Bern(\pi_j). \\ 
\end{split}
\end{equation}
When $\gamma_j=1$, the $j$th predictor is included in the regression model, when $\gamma_j=0$, the $j$th predictor is omitted from the model. Note that the indicator variable selection model can be shown to be equivalent to the spike and slab model when the spike and slab prior is imposed on $\beta_j\gamma_j $ and the spike distribution is replaced by a Dirac delta function $\delta_0(.)$ with all mass at 0. 

In many cases, predictors naturally form groups. Genetic variants located in close physical proximity and genes in the same functional pathway are all examples of groups. It is usually of interest to identify which groups of predictors are selected and which predictors are selected within each group. \cite{xu2015bayesian} and \cite{zhang2014bayesian} extended the spike and slab model to encourage shrinkage of coefficients both at the group level and within groups. For group selection, the model is largely the same as the regular spike and slab model, except that coefficients are now considered in groups so that the prior is given at the group level:
\begin{equation}
\begin{split}
\vec{\beta}_g & = \textbf{V}_g^{1/2}\vec{b}_g,\quad \textbf{V}_g^{1/2}= \mathrm{diag}(\tau_{g1},\ldots,\tau_{gm_g}); \\  
\vec{b}_g &\sim (1-\pi_1)\delta_0(.)+\pi_1 N_{m_g}(\vec{0},\textbf{I}_{m_g}); \\
\tau_{gj} & \sim (1-\pi_2)\delta_0(.) + \pi_2 N^{+}(0,s^2), 
\end{split}
\end{equation}
where $g$ is the group index and $m_g$ is the number of predictors in $g$th group. $\vec{\beta}_g=(\beta_{g1},\ldots,\beta_{gm_g})$ and $N^{+}$ denotes a normal distribution truncated below at 0. Such a model can be reparameterized as: 
\begin{equation}
\begin{split}
\beta_{gj} = \gamma_g^{(1)}\gamma_{gj}^{(2)}b_{gj}, \quad \gamma_g^{(1)} & \sim Bern(\pi_1), \quad \gamma_{gj}^{(2)}\sim Bern(\pi_2), \\ 
b_{gj} & \sim N(0,s^2),  
\end{split}
\end{equation}
which can be treated as an extension of the original indicator variable selection model to group selection as adopted by other authors \citep{chen2016bayesian, zhu2019bayesian}.

All the above models are restricted to univariate response. In applications, it is common to jointly analyze several correlated outcome measures together. Consider a multivariate regression model with $q$ correlated responses $\textbf{Y}$ of size $n\times q$ given covariate matrix $\textbf{X}$ of size $n\times p$:
\begin{equation}
\label{eq:6}
\textbf{Y} = \textbf{XB} + \boldsymbol{\epsilon} , \quad \boldsymbol{\epsilon} \sim MN_{n\times q}(\textbf{0},\textbf{I}_n \otimes  \Sigma),
\end{equation}
where $\Sigma$ is the $q\times q$ covariance matrix for the responses. Following the formulation by \cite{xu2015bayesian}, \cite{liquet2017bayesian} provides a full Bayesian multivariate spike and slab model which allows sparsity both at group level and within groups:
\begin{equation}
\label{eq:7}
\begin{split}
\textbf{B}_g & =\textbf{V}_g^{\frac{1}{2}}\Tilde{\textbf{B}}_g, \quad \textbf{V}^{\frac{1}{2}}_g=\mathrm{diag} \{ \tau_{g1},...,\tau_{gm_g} \} ;\\ 
Vec(\Tilde{\textbf{B}}_g^T|\Sigma,\tau_g,\pi_1) & \sim (1-\pi_1) \delta_0(.) + \pi_1 MN_{m_g \times q}(\textbf{0},\textbf{I}_{m_g}\otimes \Sigma); \\ 
\tau_{gj} & \sim (1-\pi_2)\delta_0(.) + \pi_2 N^+(0,s^2) , \\
\end{split}
\end{equation}
where $\textbf{B}_g$ is the $m_g\times q$ regression coefficient matrix for group $g$. In this model, both group selection and within group selection are common to all $q$ responses but not specific to any single response. 

The spike and slab and indicator variable selection models and their variants have been a popular choice for variable selection and widely used for genetic fine mapping and related applications \citep{stephens2013unified,kichaev2014integrating, hormozdiari2016colocalization,fang2018bayesian,schaid2018genome}. With univariate response, the spike and slab model and indicator selection model can be shown equivalent thus perform equally well for individual and group-level selection of predictors \citep{kuo1998variable,xu2015bayesian}. With multivariate response, \cite{liquet2017bayesian}'s multivariate spike and slab model considers predictor selection common to all responses. However, this is a rather strong assumption which does not always hold in real application. In fine mapping, heterogeneity exists for different traits, e.g. certain causal variants might target at only a subset of traits but not others. This has motivated us to propose a multivariate Bayesian variable selection method with response specific feature selection that can be applied to multi-trait fine mapping problem.


\section{A new multivariate Bayesian variable selection method for multi-trait fine mapping}
\label{sec:method}

\subsection{The model}
\label{subsec:model}

\begin{figure}[t] 
\centering
\includegraphics[scale=0.28]{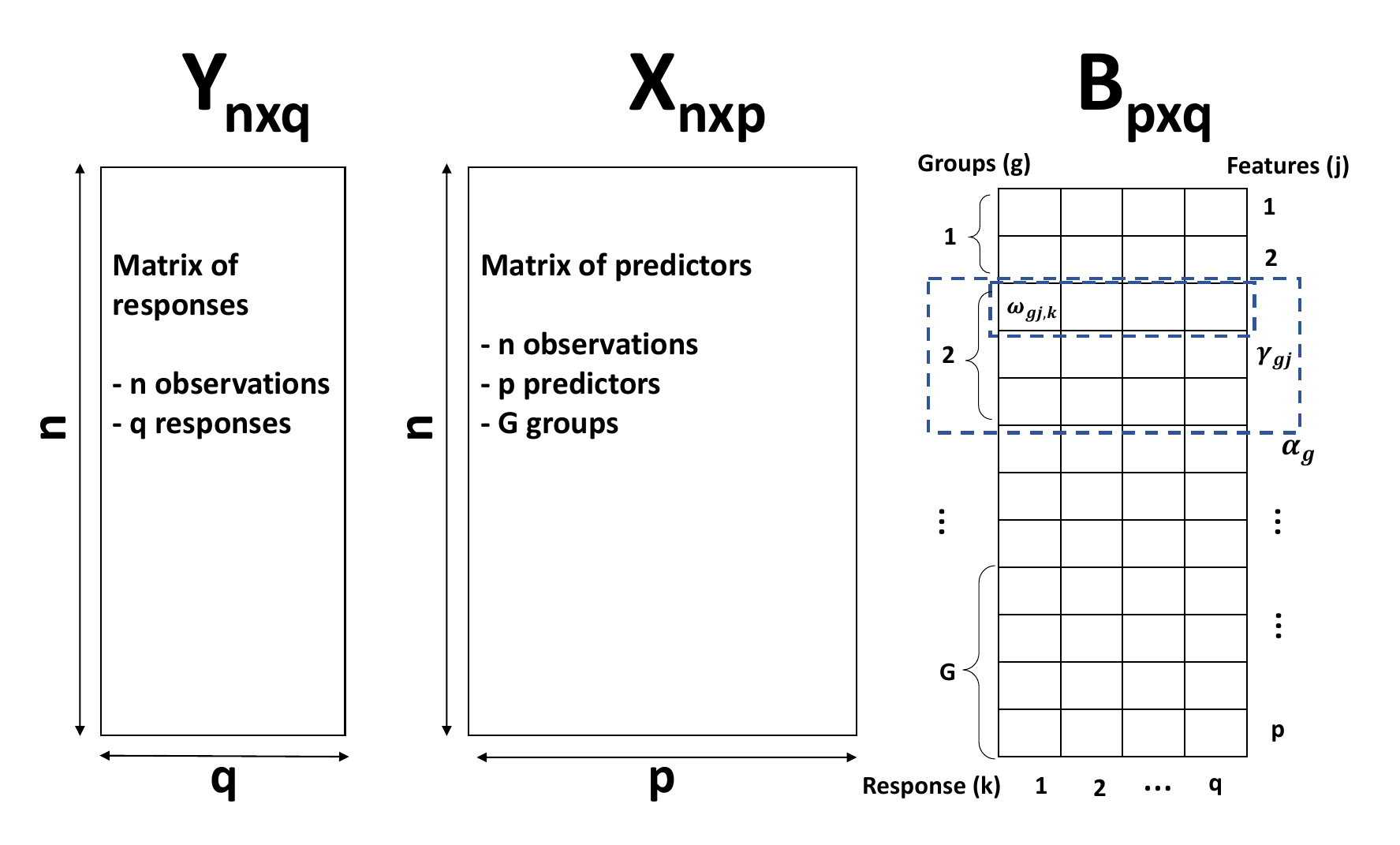}
\caption{Structure of the matrix of response \textbf{Y}, the matrix of predictor \textbf{X}, the coefficient matrix \textbf{B} in the proposed model. $\alpha_{g}$, $\gamma_{gj}$ and $\omega_{gj,k}$ correspond to the indicators for selecting the group, individual feature within the group and selecting the feature for a specific response, respectively.}
\label{fig:2}
\end{figure}

Under the same multivariate linear regression setting in Equation (\ref{eq:6}), we assume there are $p$ predictors that are potentially predictive of $q$ correlated responses. In fine mapping, this could correspond to $p$ SNPs in a local genomic region that might be causing the change in $q$ related phenotypes/traits. Denote by $\textbf{Y}$ an $n\times q$ matrix of responses and $\textbf{X}$ an $n\times p$ matrix of predictors, suppose the $p$ predictors are divided into $G$ disjoint groups (e.g. functionally annotated regions or LD blocks of SNPs in fine mapping) where $m_g$ is the number of predictors in $g$th group (i.e. $p=\sum\limits_{g=1}^G m_g $). We propose the following full Bayesian hierarchical model to encourage variable selection both at group level and within group, and specific to each response:     
\begin{equation}
\begin{split}
\label{eqn:model}
\textbf{Y} & \sim MN_{n\times q}(\textbf{XB},\textbf{I}_n \otimes  \Sigma), \quad \textbf{B} =\textbf{Z}\odot \textbf{b}, \quad \Sigma \sim \mathrm{IW}(q,\textbf{I}); \\
z_{gj,k} & = \alpha_g\gamma_{gj}\omega_{gj,k}, \quad g=1,\ldots, G; \quad j=1,\ldots,m_g; \\ 
& \qquad \qquad \qquad \qquad  k=1,\ldots,q; \\  
\vec{b}_{gj} & \sim N_q(\vec{0},s^2\Sigma), \quad g=1,\ldots, G; \quad j=1,\ldots,m_g; \\ 
\alpha_g & \sim \mathrm{Bern}(\pi^{(\alpha)}), \quad \gamma_{gj}\sim \mathrm{Bern}(\pi_{g}^{(\gamma)}), \quad \omega_{gj,k} \sim \mathrm{Bern}(\pi_{gj}^{(\omega)}), 
\end{split}
\end{equation}
where $\textbf{B}= \{(\beta_{gj,k}), g=1,\ldots, G; j=1,\ldots,m_g; k=1,\ldots,q \}$, $\textbf{Z}= \{(z_{gj,k}),g=1,\ldots, G; j=1,\ldots,m_g; k=1,\ldots,q \}$, $\textbf{b}= \{(b_{gj,k}),g=1,\ldots, G; j=1,\ldots,m_g ;k=1,\ldots,q \}$, $\vec{b}_{gj}=(b_{gj,1},\ldots,b_{gj,q})^T$, $\mathrm{IW}(q,\textbf{I})$ refers to the inverse Wishart distribution with scale matrix $\textbf{I}$ and degree of freedom $q$. Each component in $\textbf{B}$ is a product of selection indicator in $\textbf{Z}$ matrix and the underlying effect size in $\textbf{b}$ matrix, where $\odot$ represents the Hadamard (element-wise) multiplication. Each selection indicator $z_{gj,k}$ is a product of three indicator variables: $\alpha_g$ which indicates whether $g$th group of predictors is selected, $\gamma_{gj}$ which indicates whether $j$th feature in $g$th group is selected and $\omega_{gj,k}$ which indicates whether $j$th feature in $g$th group is selected specifically for $k$th response. Figure \ref{fig:2} shows the structure of the matrix of response \textbf{Y}, the matrix of predictors \textbf{X} and the coefficient matrix \textbf{B} and the correspondence of the three selection indicators to the components in \textbf{B}. Note that $z_{gj,k}=1$ only when $\alpha_g=\gamma_{gj}=\omega_{gj,k}=1$, i.e. the group that includes the predictor is selected, the predictor is selected within the group, and the selection is specific to that response. 

We assume conjugate priors on the hyperparameters: $\pi^{(\alpha)} \sim \mathrm{Beta}(1,1)$, $\pi^{(\gamma)}_{g}\sim \mathrm{Beta}(1,1)$, $\pi^{(\omega)}_{gj}\sim \mathrm{Beta}(1,1)$ and $p(s^2) \propto IG(1,v)$, where $IG(.)$ refers to inverse gamma distribution and $v$ is updated according to the Monte Carlo EM Algorithm following \cite{liquet2017bayesian} (see section 1 of the Supplement for the updating function of $v$). Note that even if each of the prior distributions $\mathrm{Beta}(1,1)$ is noninformative, the marginal prior of the products $z_{gj,k}$ is not noninformative but skewed towards zero. In fine mapping, whether we select a group of variants or an individual variant will largely depend on prior biological knowledge. For example, a variant is more likely to be causal if it plays some regulatory role, i.e. being an eQTL or multi-omics QTL. In this case, we need to impose more informative priors on the hyperparameters $\pi^{(\alpha)}$, $\pi^{(\gamma)}_{g}$ and $ \pi^{(\omega)}_{gj}$. We show one example of putting informative prior on $\pi^{(\gamma)}_{g}$ to incorporate regulatory QTL information from reference database to prioritize causal variant selection in fine mapping in this paper (see Section \ref{subsec:prior} for details). The general idea also applies to the other hyperparameters. For example, in fine mapping of different regions with different heritabilities, adaptive priors on hyperparameters as in \cite{zou2023mvsusie} might work better to reflect the complexity in real examples. Due to a different focus of the proposed method and the space limit, we do not further discuss all these possibilities.


\noindent\textbf{REMARKS:} \\
1. The three indicators play different roles in feature selection of the hierarchical model as indicated in Fig \ref{fig:2}: $\alpha_g$ controls the selection at the group level, $\gamma_{gj}$ controls the selection of individual predictor within the group, and $\omega_{gj,k}$ selects the predictor specific to each response. The product $\alpha_g\gamma_{gj}$ has the same formulation as in \cite{xu2015bayesian}'s bi-level sparse group selection with spike and slab prior, however, such a product does not discriminate the heterogeneous predictor effect on different responses. Adding $\omega_{gj,k}$ to the product will help identify which response(s) each predictor targets at. Under this setting, both the important features and the responses they target at can be identified. Note that this is very different from performing feature selection in a separate regression for each response, where no information can be shared across the different regressions. 

\noindent 2. In this model, instead of putting a multivariate spike and slab prior directly on the coefficient matrix of $\textbf{B}$ as in \cite{liquet2017bayesian}, we impose element-wise prior on each coefficient by extending \cite{kuo1998variable}'s indicator variable selection model to multivariate case. This, however, does not imply the independence between predictors or responses in the selection. The predictors belonging to the same group are more likely to be selected together since they are under the control of a common group indicator $\alpha_g$, while the same predictors are more likely to be selected for different responses due to the common individual predictor indicator $\gamma_{gj}$. 

\noindent 3. For the covariance matrix $\Sigma$, we assume it to follow a noninformative conjugate prior, the inverse Wishart distribution with degree of freedom equal to $q$ and scale matrix be the identity matrix $I$. $q$ has to be greater than 2 to ensure the
existence of $\textbf{E}(\Sigma)$. Such prior works well when we have a relatively large sample size (i.e. $n$ is large) and moderate number of responses (i.e. $q$ not too large), which holds true in genetic fine mapping. In other applications such as in transcriptomic studies, sample size is usually small, we might need to consider simultaneous variable selection and covariance selection by posing extra spike and slab or indicator variable priors on $\Sigma$ as in \cite{deshpande2019simultaneous}. The performance will need to be further investigated in future studies. 

\noindent 4. We assume the effect size vector $\vec{b}_{gj}$ has a prior multivariate normal distribution with covariance matrix $\Sigma$ scaled by a common factor $s^2$. Alternatively, we can assume a group specific $s_{g}^2$ or feature specific variance $s_{gj}^2$ to allow for more heterogeneous effect sizes. This is like the fixed effect model vs. random effect model comparison in the meta-analysis literature \citep{borenstein2010basic}. We investigated and compared the performance in simulations and did not find clear difference between the two priors. For computational efficiency, we will stick to the prior with the common scaling factor in this paper. 

\noindent 5. When $z_{gj,k}=0$, at least one of the three binary indicators $\alpha_g$, $\gamma_{gj}$ and $\omega_{gj,k}$ is equal to zero. However, the different situations (e.g. whether $\alpha_g=\gamma_{gj}=\omega_{gj,k}=0$ or only $\alpha_g=0$) are not identifiable by likelihood. This could be less of an issue since the selection is eventually determined by the product $z_{gj,k}$ specific to each response for each variable. Thus, we followed from \cite{zhu2019bayesian} and decided not to put any constraints in the prior to identify individual indicators at a price that these indicators might sometimes not be directly interpretable. 

\noindent 6. The current model focuses on a set of normally distributed continuous outcomes. Extension to binary, categorical and other types of outcomes or mixed types of outcomes is possible, though the multivariate distribution of correlated non-normal outcomes is not as well defined as normal outcomes. Alternatively, one can adopt data augmentation algorithm \citep{albert1993bayesian} by introducing latent variable for each outcome and build the hierarchy on the latent variables. For example, for binary outcome, we can introduce a latent continuous variable $T_{i,k} (i=1,\ldots,n, k=1,\ldots,q)$ to replace the binary $Y_{i,k}$ in the regression, i.e. we assume $Y_{i,k}=1$ if $T_{i,k}\ge 0$ and $Y_{i,k}=0$ otherwise, and $\textbf{T} \sim MN_{n\times q}(\textbf{XB},\textbf{I}_n \otimes  \Sigma)$. 

\noindent 7. Similar to \cite{xu2015bayesian} and \cite{liquet2017bayesian}, our method also assumes predictors naturally form groups, and such a bi-level selection has demonstrated a significant improvement in variable selection and prediction. However, the definition of groups could be subjective and context-dependent. In addition, there will be quite a few predictors which do not belong to any groups known as ``singletons''. For singletons, $\alpha_g$ and $\gamma_{gj}$ will always be equal to each other.  

\noindent 8. In practice, important covariates that might confound the selection results need to be included and carefully addressed. Since our method is built upon a very general multivariate regression model, any covariates can be readily adjusted in the model for their additive or non-addictive effects.

\subsection{Prioritizing selection using existing biological knowledge}
\label{subsec:prior}

In this subsection, we show an example of incorporating biological knowledge in the prior distribution of hyperparameters $\pi$'s in fine mapping application. eQTLs are genomic loci that explain variation in gene expression levels. Genetic studies have found SNPs associated with complex traits more likely to be eQTLs than other SNPs on genotype arrays with the same allele frequencies \citep{nicolae2010trait}. Thus, it is natural to put more weights on the eQTLs identified from large-scale genomic studies when determining causal variants in fine mapping. In our study, we incorporate eQTL information from reference database (e.g. ENCODE, ROSMAP, GTEx, QTLbase) and consider the following prior on the selection indicator $\gamma_{gj}$ to prioritize the eQTLs:
\begin{equation}
\begin{split}
\label{eq:9}
\gamma_{gj}\sim \text{Bern}(\pi^{(\gamma)}_{gj}), \quad \Phi^{-1}(\pi^{(\gamma)}_{gj})= d_0 +  d_1 A_{gj}, \quad d_0 & \sim N(0,\sigma^2), \\ 
d_1 & \sim N(\mu_d,\tau^2),
\end{split}
\end{equation}
where $A_{gj}$ is a binary variable indicating whether $j$th feature in $g$th group is an eQTL based on the reference database. $\Phi^{-1}(.)$ is the inverse of CDF of standard normal distribution, a.k.a. probit link in Bayesian literature to achieve conjugacy \citep{albert1993bayesian}. We note that there is some flexibility in setting prior for $d_1$, and in particular we may control the relative strength/emphasis of eQTL information by varying the value of the prior mean $\mu_d$. If there is a strong belief that eQTL should indicate causality, the prior mean $\mu_d$ may be set more positive value so the probability of the corresponding eQTL being a causal variant is increased. Alternatively, one can use gamma distribution truncated below zero as the prior for $d_1$ to put moderately larger weights on eQTLs. In the real data application, we set $\mu_d=0$. Lastly we allow noninformative priors on $\sigma^2$ and $\tau^2$, e.g. following Jeffreys priors. Similar formulation for incorporating functional annotation or eQTL information in fine mapping has also been seen in \cite{kichaev2017improved}. 

All the parameters in the model have closed form posterior distributions, thus we use a Gibbs sampler to sample from their posterior distributions. The detailed conditional posterior distributions of all the parameters can be found in section 1 of the Supplement.

\subsection{Posterior median as a soft thresholding estimator}
\label{subsec:theory}

It has been shown that under mild conditions, the posterior median estimator of normal mean sample with spike-and-slab prior is a soft-thresholding estimator with desired oracle (i.e. selection consistency) and asymptotic normality properties \citep{johnstone2004needles,xu2015bayesian}. \cite{liquet2017bayesian} generalized the thresholding results to multivariate response variable and \cite{zhu2019bayesian} showed the results for the equivalent reparametrized indicator variable selection model. In this paper, we will also use posterior median estimator for both variable selection and parameter estimation. In this subsection, we will show that the posterior median estimator is a soft thresholding estimator in our model and we will prove its oracle and asymptotic normality properties. 

Assume that the design matrix $\textbf{X}$ is orthogonal (that is, $\textbf{X}^T\textbf{X}=n\textbf{I}_p$), consider the model described in equation (\ref{eqn:model}) with fixed $\Sigma$ and $s^2$, it can be shown that the marginal posterior distribution of $\beta_{gj,k}$ given the observed data is a spike and slab distribution (see derivation in section 2 of the Supplement):
\begin{equation}
\beta_{gj,k}|\textbf{X},\textbf{Y}\sim l_{gj,k}N((1-D_{n})\hat{\beta}_{gj,k}^{LS},\frac{1-D_{n}}{n}\Sigma_{kk})+(1-l_{gj,k})\delta_{0}(\beta_{gj,k}),
\end{equation}
where $\hat{\beta}_{gj,k}^{LS}$ is the least squares estimator of $\beta_{gj,k}$, $D_n=\frac{1}{1+ns^2}$, $\Sigma_{kk}$ is the $k$th diagonal element of $\Sigma$ and 
\begin{equation}
\begin{split}
l_{gj,k} &=P(z_{gj,k}=1| rest) \\
& = \frac{\pi_{gj}^{*} (1+ns^2)^{\frac{1}{2}}\exp{\left[(1-D_n)n\Sigma_{kk}^{-1}(\hat{\beta}_{gj,k}^{LS})^2\right]}  }{\pi_{gj}^{*}(1+ns^2)^{\frac{1}{2}}\exp{\left[(1-D_n)n\Sigma_{kk}^{-1}(\hat{\beta}_{gj,k}^{LS})^2\right]} + (1-\pi_{gj}^{*}) } ,
\end{split}
\end{equation}
where $\pi_{gj}^*=\pi^{(\alpha)} \pi_{g}^{(\gamma)} \pi_{gj}^{(\omega)}$.

The resulting posterior median is a soft thresholding estimator $\hat{\beta}_{gj,k}^{Med}= \mathrm{Med}(\beta_{gj,k}|\textbf{X},\textbf{Y})$ given by (see the derivation in section 3 of the Supplement):

\begin{equation}
\begin{split}
\hat{\beta}_{gj,k}^{Med}= \mathrm{sgn}(\hat{\beta}_{gj,k}^{LS}) & \Bigg((1-D_n)|\hat{\beta}_{gj,k}^{LS}|-  \\ 
& \frac{\sqrt{\Sigma_{kk}}}{\sqrt{n}}\sqrt{1-D_n}\Phi^{-1}\left(\frac{1}{2\mathrm{max}\left(l_{gj,k},\frac{1}{2}\right)}\right)\Bigg)_+,
\end{split}
\end{equation}
where $\mathrm{sgn}(\cdot)$ is the sign function, $\Phi$ is the CDF of the standard normal distribution and $(x)_+$ takes the value of $x$ if $x>0$ and zero otherwise. This is similar to the lasso estimator \citep{tibshirani1996regression} which can be expressed as a soft-thresholding estimator under orthogonal design. The results also match well with that by \cite{xu2015bayesian} for univariate response and \cite{liquet2017bayesian} for multivariate response vector, except that what we derive is specific to each response in the multivariate case. 

Next, we show the asymptotic properties of the posterior median estimator $\hat{\beta}_{gj,k}^{Med}$. Let $\mathbf{B}^0$ and $\beta_{gj,k}^0$  be the true values of $\mathbf{B}$ and $\beta_{gj,k}$. Define $\mathcal{A}=(I(\beta_{gj,k}^0\neq 0),g=1,\ldots,G;j=1,\ldots,m_g;k=1,\ldots,q)$ as the indicator vector of non-zero covariates in the true model and let $\mathcal{A}_n=(I(\hat{\beta}_{gj,k}^{Med}\neq 0),g=1,\ldots,G;j=1,\ldots,m_g;k=1,\ldots,q)$ be the indicator vector of non-zero covariates from the posterior median estimator.

Under the orthogonal design, the posterior median estimator has the oracle property, i.e. variable selection consistency.

\begin{thm}[Asymptotic Properties of the Posterior Median Estimator]
\label{theorem:oracle}
Assuming an orthogonal design matrix $\textbf{X}^T\textbf{X}=n\textbf{I},\hspace{0.25em}\sqrt{n}s^2\rightarrow\infty$ and $\frac{\log(s^2)}{n}\rightarrow 0$ as $n\rightarrow \infty$, then the posterior median estimator in our model has the following asymptotic properties:
\begin{center}
$\lim_{n\rightarrow\infty}P(\mathcal{A}_n=\mathcal{A})=1$ (Selection consistency)

$\sqrt{n}(\widehat{\mathbf{B}}_\mathcal{A}^{Med}-\mathbf{B}_\mathcal{A}^0)\rightarrow MN(\mathbf{0},\mathbf{I}\otimes\Sigma)$ (Asymptotic normality)
\end{center}
\end{thm}
Detailed proof of this theorem can be found in section 4 of the Supplement.

\subsection{Posterior inference: Bayesian false discovery rate (BFDR) and subset posterior inclusion probabilities (subset PIP)}\label{subsec:fdr}

Practical implementations of Bayesian variable selection approaches typically summarize the posterior distribution by the marginal posterior inclusion probability (PIP) of each variable and draws inference \citep{kruschke2014doing}. Adopting the conjunction hypothesis setting $H_0: \underset{k}{\cap}\{\beta_{gj,k} =0 \}$ vs. $H_a:\underset{k}{\cup}\{\beta_{gj,k} \neq 0 \}$ (a.k.a. omnibus hypothesis), we can define PIP for $j$th predictor in group $g$ in our model as:
\begin{equation}
PIP_{gj} \coloneqq P_{gj}(H_a|\textbf{X},\textbf{Y} )= P(\underset{k}{\cup}\{\beta_{gj,k} \neq 0 \}| \textbf{X},\textbf{Y} ).
\end{equation}

Our goal is to find a list of predictors for which $H_a$ is true, while ensuring the individual level false discovery rate (FDR) is under control. We followed \cite{newton2004bfdr} to use a direct posterior probability approach. We ranked the predictors by increasing values of $P_{gj}(H_0|\textbf{X},\textbf{Y})=P(\underset{k}{\cap}\{\beta_{gj,k} =0 \} |\textbf{X},\textbf{Y})= 1-PIP_{gj}$ and the final list contains predictors with $P_{gj}(H_0|\textbf{X},\textbf{Y})$ less than some boundary $t$. Given the data, the expected number of false discoveries is $\sum_{g=1}^{G}\sum_{j=1}^{m_g}P_{gj}(H_0|\textbf{X},\textbf{Y})d_{gj}(t)$, where $d_{gj}(t)=I\{P_{gj}(H_0|\textbf{X},\textbf{Y})<t\}$. We can thus define the Bayesian false discovery rate (BFDR) for any boundary $t$ as:
\begin{equation}
BFDR(t)=\frac{\sum_{g=1}^{G}\sum_{j=1}^{m_g}P_{gj}(H_0|\textbf{X},\textbf{Y})d_{gj}(t)}{\sum_{g=1}^{G}\sum_{j=1}^{m_g}d_{gj}(t)}.
\end{equation}
BFDR can be controlled at a certain $\alpha$ level by tuning $t$. 

We can use the BFDR defined above to select the top predictors. However, inference by controlling the individual predictor level BFDR is not sufficient for our multiple predictor to multiple response problem. The rejection of the conjunction of null hypotheses is often too general to be scientifically meaningful, i.e. we only know the predictor is associated with at least one response but do not know which ones. In reality, heterogeneity usually exists among the different responses. For example, in genetic fine mapping of regional neuroimaging phenotypes, we are interested in knowing what subset of regional traits that each causal variant targets at, to generate new hypotheses. To facilitate scientific finding and interpretability, after selecting the top predictors using BFDR, we further propose a novel concept of subset PIP to help select the best subset of responses each of the selected predictor targets at.

Denote by $\mathcal{M_S}\subset \{1,2,\ldots,q \}$ a subset of response indices a predictor targets at, we define the following subset PIP for each predictor, $1\le j\le m_g^\alpha, 1\le g\le G^\alpha$:
\begin{equation}
PIP_{gj}^{\mathcal{M_S}}=P(\beta_{gj,k}\neq 0,\hspace{0.25em}\text{for}\hspace{0.25em}k\in\mathcal{M_S}|\textbf{X},\textbf{Y}), 
\label{eq:subsetPIP}
\end{equation} 
where $G^\alpha$ is the remaining number of groups and $m_g^\alpha$ the remaining number of predictors in $g$th group after selecting the top predictors by controlling BFDR at $\alpha$ level.

For each predictor, we can calculate $PIP_{gj}^{\mathcal{M_S}}$ for all possible $2^q-1$ subsets and identify the best subset as the one that maximizes $PIP_{gj}^{\mathcal{M_S}}$ (i.e. $\mathcal{M_S}_{gj}^*=\mathrm{argmax}_{\mathcal{M_S}} PIP_{gj}^{\mathcal{M_S}}$). However, size bias exists wherein smaller subset tends to have higher subset PIP. For example, a predictor might target at three responses with stronger effects on two responses but relatively weaker effects on the third one. Using the subset PIP directly calculated from equation (\ref{eq:subsetPIP}) will favor the selection of the first two responses though the underlying truth is to select all the three responses. To resolve this issue, we propose a permutation based method to generate an empirical reference distribution of $PIP_{gj}^{\mathcal{M_S}}$ with the same subset size for fair comparison and best subset selection.

We randomly shuffle all elements in the estimated $\widehat{\textbf{B}}$ matrix for a total of $R$ times, so we generate $R$ matrices of $\widehat{\textbf{B}}^r, r=1,\ldots,R$. For each permuted matrix $\widehat{\textbf{B}}^r$, we recalculate $PIP_{gj}^{\mathcal{M_S},r}$ for each subset $\mathcal{M_S}$. We then summarize all the permutations into a subset specific Z-score: $Z_{gj}^{\mathcal{M_S}}=\frac{PIP_{gj}^{\mathcal{M_S}}-\frac{1}{R}\sum_{r=1}^{R}PIP_{gj}^{\mathcal{M_S},r}}{\sqrt{\frac{1}{R}\sum_{r=1}^{R}(PIP_{gj}^{\mathcal{M_S},r}-\left(\frac{1}{R}\sum_{r=1}^{R}PIP_{gj}^{\mathcal{M_S},r}\right)})^2 }.$

The best subset of responses each predictor targets at can be identified by maximizing the corresponding Z-score, i.e. $\mathcal{M_S}_{gj}^*=\mathrm{argmax}_{\mathcal{M_S}} Z_{gj}^{\mathcal{M_S}}$. Since the Z-scores are calculated by comparing to the reference distribution of the same subset size, the size bias is mitigated. The method is summarized below in Algorithm \ref{alg:perm}.

\begin{algorithm}[t]
\caption{Permutation method for best subset selection}
\label{alg:perm}
\begin{algorithmic}[1]
\State{\textit{Input:} posterior median estimate of coefficient matrix $\widehat{\textbf{B}}$} 
\State{\textit{Output:} The best response subset $\mathcal{M_S}_{gj}^*$ for each feature, $1\le j\le m_g^\alpha, 1\le g\le G^\alpha$ }
\Statex
  \For{$g \gets 1$ to $G^\alpha$}                
    \For{$j \gets 1$ to $m_g^\alpha$} 
      \For{$\mathcal{M_S}\subset \{ 1,2,\ldots,q \}$} 
      \State Calculate $PIP_{gj}^{\mathcal{M_S}}=P(\hat{\beta}_{gj,k}\neq 0,\hspace{0.25em}\text{for}\hspace{0.25em}k\in\mathcal{M_S}|\textbf{X},\textbf{Y})$, where $\hat{\beta}_{gj,k}$ is the (gj,k)th element in $\widehat{\textbf{B}}$ 
    \EndFor
  \EndFor
 \EndFor 

   \For{$r \gets 1$ to $R$}                
      \State Obtain a permuted matrix $\widehat{\textbf{B}}^r$ by randomly shuffling all elements in $\widehat{\textbf{B}}$
    \EndFor

  \For{$g \gets 1$ to $G^\alpha$}                
    \For{$j \gets 1$ to $m_g^\alpha$} 
      \For{$\mathcal{M_S}\subset \{ 1,2,\ldots,q \}$}  
      \State Calculate 
$$ Z_{gj}^{\mathcal{M_S}}=\frac{PIP_{gj}^{\mathcal{M_S}}-\frac{1}{R}\sum_{r=1}^{R}PIP_{gj}^{\mathcal{M_S},r}}{\sqrt{\frac{1}{R}\sum_{r=1}^{R}(PIP_{gj}^{\mathcal{M_S},r}-\left(\frac{1}{R}\sum_{r=1}^{R}PIP_{gj}^{\mathcal{M_S},r}\right)})^2} ,$$ 
where $PIP_{gj}^{\mathcal{M_S},r}$ is obtained the same way as $PIP_{gj}^{\mathcal{M_S}}$ but from $\widehat{\textbf{B}}^r$
    \EndFor
\State Obtain $\mathcal{M_S}_{gj}^*=\mathrm{argmax}_{\mathcal{M_S}} Z_{gj}^{\mathcal{M_S}}$ for each predictor
  \EndFor
 \EndFor

    \State \Return {$\mathcal{M_S}_{gj}^*$}
\end{algorithmic}
\end{algorithm}

\noindent\textbf{REMARK:} \\
In this paper, we show PIP defined at individual predictor level. With the inclusion of indicators at three different levels in our model, it is flexible to also define the PIP at the group level to select groups of associated predictors. Since it is not the focus of our method, we will not further discuss it here and details will be left to future studies. 

\section{Other related methods}
\label{sec:related_methods}

Many multivariate Bayesian and regularization methods have been developed to perform individual-level or group-level variable selection with multiple related responses for general application. Here, we summarize the main features and limitations of several related methods as compared to our method (namely multivariate Bayesian variable Selection for Heterogeneous responses (mvBaSH)) below and in Table 1. In addition, we also include popular fine-mapping methods (mainly Bayesian model based) that can handle multiple traits for comparison.

\begin{itemize}
    \item Multivariate Bayesian variable selection methods:
    \begin{itemize}
        \item Multvariate Bayesian Group Lasso with Spike-and-Slab Prior (MBGL-SS) \citep{liquet2017bayesian}: This multivariate model is based on a Bayesian group lasso model with an independent spike-and-slab prior for each group variable for group-level features selection, however, this model does not select features within the groups, and all grouped features selected are common to all responses.
        \item Multivariate Bayesian Sparse Group Spike-and-Slab (MBSGS) \citep{liquet2017bayesian}: This is a second model proposed in \cite{liquet2017bayesian} and a multivariate implementation of the spike-and-slab model as given in Equations \ref{eq:6} and \ref{eq:7}, which allows sparsity both at group levels and within groups. However, the selected features or groups are still common to all responses thus the heterogeneity among the responses cannot be accounted for.
        \item Multivariate Sum of Single Effects Regression (mvSuSIE) \citep{zou2023mvsusie}: This is a multivariate extension of the SuSIE model \citep{wang2020simple}, which is a new formulation of Bayesian variable selection regression as an alternative to spike and slab prior. This method is also inspired by fine-mapping, but applicable to a wider range of applications. The mvSuSIE model selects variables simultaneously for multiple outcomes and is able to select specific outcomes associated with a variable by using a trait-specific measure local false sign rate (lfsr).
     \end{itemize}

 \item Multivariate regularization methods:
    \begin{itemize}
        \item REgularized Multivariate regression for identifying MAster Predictors (remMap) \citep{peng2010regularized}: This method adopted a mixed L1/L2 penalty to encourage the selection of ``master'' predictors that affect many response variables in multivariate linear regression. This method cannot incorporate any prior knowledge e.g. for fine mapping application, in addition, it cannot select features or groups specific to each response. 
        \item Multivariate sparse group lasso (MSGLasso): This method imposes a sparse group lasso like (L1+L2) penalty in a multivariate regression model setting with arbitrary grouping structure \citep{li2015multivariate}. Same as remMap, this method cannot incorporate any prior information, nor perform response specific variable selection.
    \end{itemize}
    \item Fine-mapping methods: 
    \begin{itemize}
        \item Causal Variants Identification in Associated Regions (CAVIAR) \citep{hormozdiari2014caviar}: This Bayesian fine-mapping method directly models the LD structure in a genomic region to identify causal variants of interest. CAVAIAR has a constraint on the maximum number of causal variants that can be detected. It is designed for a single trait, but has been applied to multiple trait setting using average PIP.
        \item Probalistic Annotation Integrator (PAINTOR) \citep{kichaev2017improved}: This is a multivariate extension of the original Bayesian fine-mapping model for single trait \citep{kichaev2014integrating} to target at multiple traits. PAINTOR directly models LD structure and is able to incorporate additional functional annotation data. Same as CAVIAR, PAINTOR also has a constraint of the maximum number of causal variants one can select.
        \item mvBIMBAM \citep{stephens2013unified}: mvBIMBAM is a general framework to establish both direct and indirect associations between SNPs and a set of correlated outcomes. The main purpose of the model is to reveal the causal relationship between genetics and multiple phenotypes, so it does not perform well when there are many phenotypes (the possible number of causal relationships goes exponentially large as the number of outcomes increases). One other downside of this model is instead of using PIP that has better probabilistic interpretation, it uses Bayes factor for inference. 
    \end{itemize}
\end{itemize}

\section{Simulation}
\label{sec:simulation}

\subsection{Simulation setting}

In this section, we conducted simulation studies to evaluate the variable selection and prediction performance of our method. Since the primary motivation for our method is genetic fine mapping, we simulated both genotype ($\textbf{X}$) and multivariate phenotype data ($\textbf{Y}$) to reflect the fine mapping application. 


We started by simulating the genotype data $\textbf{X}$. To mimic the real genotype pattern, we followed from \cite{wang2007improving,han2012Latent} and simulate the number of minor alleles for all SNPs with fixed minor allele frequency of 0.24 using latent Gaussian model with covariance matrix defined by the LD structure. To mimic the grouping and correlation structure in genotype data, we simulated LD blocks of SNPs where SNPs within the same block has fixed correlation $\rho_\text{within}=0.6$ and SNPs not in the same block has correlation $\rho_\text{between}=0.3$. We allow group sizes to vary, belonging to the set $\{5,10,20\}$.


Next we simulated the coefficient matrix \textbf{B} and the outcome $\textbf{Y}$. We assumed $q=6$ correlated traits and the outcome $\textbf{Y}_i$ of each sample is simulated from $MN(\textbf{X}_i \textbf{B},\Sigma), i=1,\ldots,n$, where $\Sigma$ is the covariance matrix assuming to have a compound symmetry structure with common variance $\sigma^2_Y$ and fixed correlation $\rho_Y=0.66$.
 
To simulate the coefficient matrix \textbf{B} (which is a product of the selection indicator matrix \textbf{Z} and the effect size matrix \textbf{b}), we designed five simulation scenarios as below for a comprehensive assessment of our method, by varying the dimensionality (i.e. whether the total number of variants is greater or smaller than the sample size), the level of heterogeneity of response (i.e. homogeneous where the causal variants target at all responses alike vs heterogeneous where the causal variants only target at subset of responses), and the distributions of signals (i.e. whether the causal variants are evenly distributed or clustered within groups). We also considered two heritability levels \citep{zhu2020heritability},  i.e. the proportion of overall variance in phenotype that can be explained by the causal variants: $h^2=7\%$ (when $s^2=0.015$, see Model in Equation \ref{eqn:model}) and $h^2=1\%$ (when $s^2=0.00175$) for each of the simulated scenarios to well mimic the real fine mapping application.

\noindent\textbf{Scenario I: Low dimension, homogeneous responses, signals evenly distributed.} We considered $n=500$ samples and $p=100$ SNPs, among them $c=5$ SNPs are true causal variants and are evenly distributed over all SNPs. We assumed $\sigma^2_Y=10$. For the true causal variants, we assumed $\vec{\beta}_{gj} \sim N(\vec{0},s^2\Sigma) $. For non-causal SNPs, we assumed $\beta_{gj,k}=0$ for $1\le k\le 6$.

\noindent\textbf{Scenario II: Low dimension, homogeneous responses, signals clustered within groups.} We considered the same $n$, $p$, $\sigma^2_Y$ and the same number of true causal variants $c$ as in scenario I but now we assumed the causal SNPs were clustered in the same group. $\beta_{gj,k}$'s were simulated the same way as in scenario I.

\noindent\textbf{Scenario III: Low dimension, heterogeneous responses, signals evenly distributed.} We considered the same $n$, $p$, $c$ and distribution of signals as in scenario I, but instead of assuming the effects of causal variants to be the same across all traits, we now considered more heterogeneous case where different causal SNPs can target at different subsets of traits. When a SNP is causal for a subset of traits (e.g. denote the subset indices by $\mathcal{M_S}$), we assumed $\vec{\beta}_{gj} \sim N(\vec{0},s^2\Sigma)$ and then force $\beta_{gj,k}=0$ for $k \notin \mathcal{M_S} $. For non-causal SNPs, we assumed $\beta_{gj,k}=0$ for $1\le k\le 6$. In heterogeneous case, we assumed $\sigma^2_Y=8$.

\noindent\textbf{Scenario IV: Low dimension, heterogeneous responses, signals clustered within groups.} The values of $n$, $p$, $c$, $\sigma^2_Y$ and the way $\beta_{gj,k}$'s were simulated remained the same as scenario III, except now the causal SNPs were clustered in the same group.

\noindent\textbf{Scenario V: High dimension, heterogeneous responses, signals clustered within groups.} We considered the same setting as in scenario IV, except now we considered a high-dimensional case with $n=300$ and $p=500$.



Each simulation was replicated for 100 times. To benchmark our method, we compared to the other multivariate Bayesian variable selection methods including Multivariate Bayesian Sparse Group Spike-and-Slab (MBSGS) \citep{liquet2017bayesian} and Multivariate Sum of Single Effects Regression (mvSuSiE) \citep{wang2020simple, zou2023mvsusie} using both overall PIP and trait-specific lfsr to rank the SNPs, the multivariate regularization methods including remMAP \citep{peng2010regularized} and MSGLasso \citep{li2015multivariate}, as well as the popular Bayesian fine mapping method CAVIAR \citep{hormozdiari2014caviar} and PAINTOR \citep{kichaev2017improved}. Although mvBIMBAM is a multivariate method that may be used in genetic fine mapping applications, we chose not to include it in our comparisons because of its underlying goal to detect causal relationship among different responses which differs from the main goal of fine mapping. We assessed both the variable selection and prediction performance of each method. For each method, we sorted the features by the main output measure for inference (e.g. PIP or lfsr for mvSuSiE). To eliminate the influence of arbitrary cutoffs in different methods, we derived sensitivity and specificity of variable selection under different cutoffs and calculated the area under the receiver operating curves (AUC) for a fair comparison. To evaluate prediction performance, we estimated the coefficients for all regression model based methods (mvBaSH, MBSGS, mvSuSiE, remMAP and MSGLasso) and used a five-fold cross-validation to calculate Mean Square Prediction Error (MSPE), defined as the average of $\text{MSPE}=\frac{1}{q\cdot n_{\text{test}}}\sum_{i=1}^{n_{\text{test}}}\sum_{k=1}^{q}\left((\textbf{X}^{\text{test}}\hat{\textbf{B}})_{i,k}-\textbf{Y}^{\text{test}}_{i,k}\right)^2$ over all five folds, where $n_{\text{test}}$ is the sample size in the testing set. In addition, we also assessed how the actual false discovery rate (FDR; the number of false positives/the number of claimed positives) and false
omission rates (FOR; the number of false negatives/number of claimed negatives) were controlled at a nominal FDR level of 10\%.

R was used to run all methods. For both our method and MBSGS, in Scenarios (I)-(IV) we ran for 10000 MCMC iterations with the first 7500 iterations as burn-in. To calculate sensitivity and specificity for remMap and MSGLasso, a variety of tuning parameters were considered. 


\subsection{Simulation results}

Table \ref{tab:simulations} showed the variable selection and prediction performance for all methods in the five scenarios when $h^2=7\%$. When the responses were homogeneous and signals evenly distributed (Scenario I), MBSGS, mvSuSIE and our method performed almost equally well and outperformed other fine mapping and regularization methods in variable selection (higher AUC, FDR well controlled). When the signals were clustered (Scenario II), our method and MBSGS which incorporated the group information had improved performance and outperformed all other methods (higher AUC). When the responses became more heterogeneous (Scenario III-IV), our mvBaSH method, which selects features specific to each response, outperformed MBSGS and the other variable selection and fine mapping methods (higher AUC and better controlled FDR). This is partly because MBSGS looked at the multivariate response as a whole thus were less sensitive to heterogeneous signals especially when the signal strength was weak, while our method conducted response-specific selection thus can capture heterogeneous and weak signals. Note that increasing the dimension of the parameter space (Scenario V) impacted the performance of all methods as expected, but our method remained the best performer. When $h^2=1\%$, the performance of all methods decreased but our method was still advantageous than other methods when signals were clustered, responses were heterogeneous and in high-dimensional case (see Table S1). For computational cost, as both our method and MBSGS \citep{liquet2017bayesian} are MCMC sampling base, they are generally slower than variational approximation based mvSuSIE method \citep{zou2023mvsusie}. Though other regularization and fine mapping methods are faster than us in low dimensional case, they turn to be unstable and slow in convergence in high dimensional case (see Table S2). Considering our advantage in improved power and accuracy, it is worth the relatively longer computational time.


\section{Real data application}
\label{sec:application}

In this section, we applied our method to a real multi-trait fine mapping problem to identify the causal variants of several related addictive behaviors and their associated risk factors using data from the UK BioBank (UKB) cohort. UKB is a large UK based biomedical database that collected comprehensive genetic and phenotypic details of over $500,000$ participants aged 40-69 recruited in the years 2006-2010 \citep{sudlow2015ukbiobank}.  Recent studies have reported moderate to high genetic correlation among the addictive behaviors such as smoking, alcohol and cannabis use \citep{quach2020expanding,saunders2022genetic}. In this study, we considered four addictive behaviors including cigarettes per day (CPD), age of smoking initiation, alcohol use, cannabis use and one common genetically related risk factor BMI \citep{wills2017phenotypic,wills2019phenotypic} as the responses ($q=5$). For genotype data, we followed common genotype encoding by counting the number of minor alleles and represented them as 0, 1 and 2 to measure the additive effect at a SNP. We focused on participants of European origin and kept only participants with complete data on all the responses, and further excluded those participants who had CVD, hyptertension, obesity, diabetes, mental or behavioral issues, diseases of the nervous system, and disease of the cardiovascular system, giving rise to a final sample size of $n=3726$. 

\begin{figure*}[!h]
\centering
\includegraphics[scale=0.45]{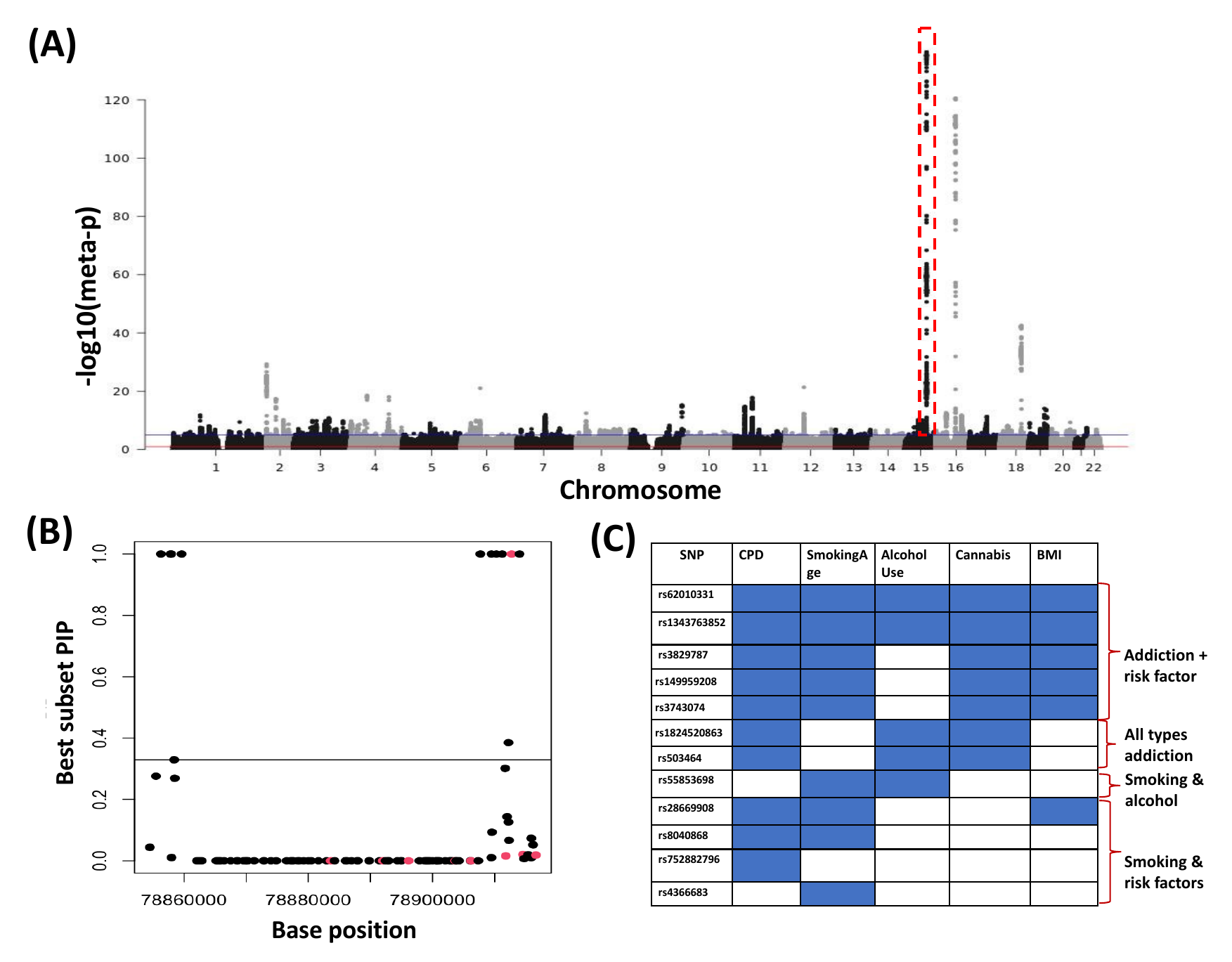}
\caption{Real data analysis results. (A) Manhattan plot of meta-analyzed p-value from GWAS on each trait. (B) Selection results by our method, y-axis is the best subset PIP, x-axis is the base position and the line denotes the cutoff BFDR$<0.1$ for selecting the causal variants, eQTLs are highlighted in red. (C) The best subset of traits each causal variant targets at, blue box indicates that the corresponding response is selected for that SNP.}
\label{fig:real}
\end{figure*}

We first performed GWAS on each trait separately and meta-analyzed the GWAS summary results using adaptively weighted Fisher's method \citep{li2011adaptively}. Shown on the Manhattan plot using meta-analyzed p-value, a genomic region on chromosome 15 was found highly associated with multiple addiction related traits (highlighted in Fig \ref{fig:real}A), so we zoomed into that region with 50k bp extension both upstream and downstream. The region includes a set of 107 SNPs ($p=107$) for multi-trait fine mapping analysis to identify the causal variants. 

 We grouped the SNPs by their annotation categories in GENCODE \citep{harrow2012gencode} (e.g. intron, exon, intergenic) and the genes in proximity. Many causal variants act synergistically to impact phenotype \citep{corradin2014combinatorial}. Grouping by these annotation categories will help improve the detection power and select causal variants on the basis of their related functional roles. One can also use an online annotator like FAVOR to assign groups based on the plausible functional consequences of variants  \citep{zhou2023favor}. In addition, we incorporated eQTL information from GTEx database \citep{gtex2015genotype} to prioritize the selection of causal variants using existing biological knowledge. Our mvBaSH method identified 12 causal SNPs at BFDR$<0.1$ (see Fig \ref{fig:real}B), among them one is an eQTL (``rs149959208''). Using subset PIP, These variants targeted at different subsets of addictive traits and risk factors (see Fig \ref{fig:real}C). For example, rs62010331 and rs1343763852 are causal to all the five traits; rs1824520863 and rs503464 target at the three major types of addiction (nicotine, alcohol and cannabis addiction);  while rs28669908 and rs8040868 are targeting at more nicotine addiction related traits (CPD, smoking age and BMI). One SNP rs503464, found to be causal to CPD, alcohol and Cannabis use, has been previous reported in multiple studies as related to smoking and alcohol drinking habits \citep{bierut2008variants,wang2009genetic}. All these SNPs are located in the genes \textit{CHRNA3} and \textit{CHRNA5} (see Table \ref{tab:real}), well-known genes that encodes for subunits of nicotinic acetylcholine receptors (nAchR) and are found related to Tobacco, Alcohol and Cannabis use and their comorbidities with reproducible findings in multiple large-scale genetic studies \citep{chen2012nicotine,erzurumluoglu2020meta,liu2019association}. Interestingly, a majority of the identified causal variants are distributed in noncoding regions including intronic and 5'UTR (see Table \ref{tab:real}), implying the potential of causal variants playing regulatory roles in the process rather than directly encoding for genes \citep{cano2020gwas}. Further analysis from SNP to gene/RNA level is needed to fully understand the underlying causal regulatory mechanism. We also ran our method when the functional group information of SNPs were not given, the method became less powerful (fewer significant SNPs detected) and the detected SNPs were scattered across the region without unifying functional theme (results not shown), affirming the importance of including group information in fine mapping. In addition to our method, we also analyzed the data using MBSGS, mvSuSIE and PAINTOR (see FigS1 for a Venn Diagram of the number of detected variants by each method). While our method performs more similar to MBSGS and mvSuSIE (4 variants overlapped) due to their Bayesian variable selection model nature, our method is more powerful and provides the subset of traits each detected causal variant targets at for further investigation. The PAINTOR method detects 6 variants but has the least overlap with the other methods. 
 
\section{Discussion}
\label{sec:discussion}

In this paper, we proposed a novel multivariate Bayesian variable selection method that selects predictors at group level, within groups as well as specific to each response for multiple correlated responses with possible heterogeneity. We also incorporated biological knowledge of eQTL information from existing database into our model to prioritize the causal variant selection in the context of fine mapping. In addition, we proposed to use BFDR to select the top predictors followed by subset PIP to identify the best subset of responses each predictor targets at to facilitate biological findings and generate new hypotheses. Our new method provided deeper biological insights into the current multi-trait fine mapping problem by integrating annotation groups and existing biological knowledge to select more functionally related variants, and characterizing the ways in which the different causal variants affect different but highly related traits that can potentially illuminate the genetics of complex diseases or traits. Under mild conditions, we showed that the posterior median estimators of regression coefficients in our model possess the selection consistency and asymptotic normality. Via extensive simulation and a real data application to fine mapping of causal variants for multiple addictive behaviors, we showed a remarkable improvement of variable selection performance of our method as compared to other multivariate Bayesian variable selection, regularization and fine-mapping methods. 

Nowadays, it is frequent for researchers to collect high-dimensional data on multiple related responses with the aim of identifying the common predictors jointly predictive of these responses from a large pool of candidates. A majority of modern variable selection methods, including both Bayesian variable selection and regularization methods, mainly target at a single response. Running variable selection for each response separately tends to ignore the correlation among responses and is less powerful. The very few options of methods that target at multivariate responses \citep{liquet2017bayesian,peng2010regularized,li2015multivariate} treat all responses as a whole without considering the heterogeneity across the responses, where each predictor might target at a different subset of responses. This is fairly common phenomenon when we jointly analyze multiple responses from one resource or from multiple studies \citep{ma2020variable}. Our method is the first Bayesian variable selection method that targets at the multivariate response heterogeneity issue by selecting predictors at multiple levels: group-level, individual feature level and response level, coupled with the best subset of response selection for each predictor. Built on the basis of the more flexible indicator variable selection, our method is easily extensible to handle more complex multi-to-multi problems, e.g. when both predictors and responses have structural patterns. In addition, it is also flexible to put priors on the hyperparameters of selection probabilities in our model to favor selection of certain predictors or certain predictor-response pairs based on existing knowledge, e.g. regulation of gene by miRNA based on motif match from existing sequence database. 

In its current implementation and application, our method only considers a relatively fewer number of responses thus we assume a noninformative inverse-wishart prior on the covariance matrix. In the case with a large number of responses with the same level as the number of predictors \citep{ke2022high}, priors to encourage simultaneous selection of regression coefficients and covariance terms are strongly recommended \citep{deshpande2019simultaneous,li2021joint}. 

We built our model on individual level (raw) genotype data, rather than GWAS summary statistics as in many fine mapping methods. This allows for a more generalizable model that may be applied to many different applications outside fine mapping. Our current application focuses on continuous outcomes, but the model is readily extensible to analyze outcomes of other data types, such as binary, count or mixed data types. Detailed performance will be evaluated in future studies.  

The computing time for the real data example of fine mapping for multiple addictive behaviors is 30 mins for 10,000 MCMC iterations with 16 CPU cores, 2.7 GHz and 128 GB RAM. Optimization of code in C++ and applying further parallel computing will likely further reduce computing time \citep{scott2016bayes,biswas2022scalable}. An efficient R package called "mvBaSH" and data used in this study are available on github at \url{https://github.com/tacanida/mvBaSH}.

\section{Tables}\label{sec5}

\begin{table*}[h]
\centering
\resizebox{\linewidth}{!}{
\small
\begin{tabular}{|c c c c c c c c|} 
 \hline
 \textbf{Method} & \textbf{Indi. feature} & \textbf{Group feature} & \textbf{Resp. specific} & \textbf{Best resp.}  & \textbf{Prior biol.} & \textbf{Constraint} & \textbf{Reference} \\  
   & \textbf{selection} & \textbf{selection} & \textbf{feature selection} & \textbf{subset} & \textbf{knowledge} & \textbf{free} & \\  
 \hline\hline
 MBGL-SS & N & Y & N & N & N & Y & \cite{liquet2017bayesian} \\
 MBSGS & Y & Y & N & N & N & Y & \cite{liquet2017bayesian} \\
 mvSuSIE & Y & Y & Y & N & Y & Y & \cite{wang2020simple} \\ 
remMap & Y & Y & N & N & N & Y & \cite{peng2010regularized} \\
MSGLasso & Y & Y & N & N & N & Y & \cite{li2015multivariate} \\
 CAVIAR & Y & N & N & N & Y & N & \cite{hormozdiari2014caviar} \\ 
 PAINTOR & Y & N & N & N & Y & N & \cite{kichaev2017improved} \\
 mvBIMBAM & Y & N & N & N & Y & Y & \cite{stephens2013unified} \\
  mvBaSH  & Y & Y & Y & Y & Y & Y & \textcolor{blue}{Our method} \\ 
 \hline
\end{tabular}}
\caption{A comparison of different multivariate variable selection methods. Indi. feature selection: whether conducting individual level feature selection; Group feature selection: whether conducting group-level feature selection; Resp. specific feature selection: whether selecting features specific to each response; Best resp. subset: whether selecting the best subset of responses each feature targets at; Prior biol. knowledge: whether prior biological knowledge can be incorporated into the model; Constraint free: whether there is any constraint in software implementation.}\label{tab:comparison}
\end{table*}

\begin{table}[H]
\centering
\small
\begin{tabular}{|p{0.15cm} p{2cm} p{2cm} p{2cm} p{2cm} p{2cm} |} 
 \hline
 \textbf{} & \textbf{Model} & \textbf{AUC} & \textbf{FDR} & \textbf{FOR} & \textbf{MSPE} \\    
 \hline\hline
I & mvBaSH & 0.92 (0.01) & 0.08 (0.02) & 0.04 (0.00) & 10.14 (0.05) \\ 
  & MBSGS & 0.93 (0.01) & 0.09 (0.02) & 0.04 (0.00) & 10.06 (0.05) \\
  & mvSuSiE$^1$ & 0.89 (0.01)  & 0.07 (0.02) & 0.04 (0.00) & 10.09 (0.05) \\
  & mvSuSiE$^2$ & 0.89 (0.01) & 0.12 (0.03) & 0.04 (0.00) & *** \\
  & CAVIAR & 0.84 (0.01) & 0.28 (0.04) & 0.04 (0.00) & - \\
  & PAINTOR & 0.63 (0.02) & 0.13 (0.03) & 0.04 (0.00) & - \\
  & remMAP & 0.60 (0.01) & - & - & 9.24 (0.04) \\
  & MSGLasso & 0.53 (0.00) & - & - & 10.13 (0.05) \\
II & mvBaSH & 1.00 (0.00) & 0.05 (0.02) & 0.04 (0.00) & 10.18 (0.05) \\
  & MBSGS & 1.00 (0.00) & 0.01 (0.01) & 0.03 (0.00) & 10.09 (0.04) \\
  & mvSuSiE$^1$ & 0.90 (0.01) & 0.05 (0.02) & 0.04 (0.00) & 10.14 (0.04) \\
  & mvSuSiE$^2$ & 0.87 (0.01) & 0.04 (0.02) & 0.04 (0.00) & *** \\
  & CAVIAR & 0.89 (0.01) & 0.19 (0.04) & 0.04 (0.00) & - \\
  & PAINTOR & 0.65 (0.02) & 0.16 (0.03) & 0.04 (0.00) & - \\
  & remMAP & 0.60 (0.01) & - & - & 9.23 (0.04) \\
  & MSGLasso & 0.53 (0.00) & - & - & 10.09 (0.05) \\
III & mvBaSH & 0.91 (0.01) & 0.13 (0.03) & 0.04 (0.00) & 7.95 (0.04) \\
  & MBSGS & 0.86 (0.01) & 0.14 (0.03) & 0.04 (0.00) & 7.94 (0.04) \\
  & mvSuSiE$^1$ & 0.81 (0.01) & 0.14 (0.03) & 0.04 (0.00) & 7.94 (0.04) \\
  & mvSuSiE$^2$ & 0.81 (0.01) & 0.13 (0.03) & 0.04 (0.00) & *** \\
  & CAVIAR & 0.74 (0.01) & 0.52 (0.05) & 0.04 (0.00) & - \\
  & PAINTOR & 0.56 (0.02) & 0.11 (0.03) & 0.05 (0.00) & - \\
  & remMAP & 0.60 (0.01) & - & - & 7.36 (0.04) \\ 
  & MSGLasso & 0.52 (0.00) & - & - & 8.00 (0.04) \\
IV & mvBaSH & 0.99 (0.00) & 0.07 (0.02) & 0.04 (0.00) & 7.97 (0.04) \\
  & MBSGS & 0.98 (0.00) & 0.08 (0.03) & 0.03 (0.00) & 7.97 (0.04) \\
  & mvSuSiE$^1$ & 0.88 (0.01)  & 0.11 (0.03) & 0.04 (0.00) & 7.98 (0.04) \\
  & mvSuSiE$^2$ & 0.82 (0.01) & 0.06 (0.02) & 0.04 (0.00) & *** \\
  & CAVIAR & 0.80 (0.01) & 0.39 (0.05) & 0.04 (0.00) & - \\
  & PAINTOR & 0.56 (0.01) & 0.11 (0.03) & 0.05 (0.00) & - \\
  & remMAP & 0.60 (0.01) & - & - & 7.30 (0.03) \\
  & MSGLasso & 0.53 (0.00) & - & - & 8.00 (0.03) \\
V & mvBaSH & 0.95 (0.02) & 0.31 (0.07) & 0.01 (0.00) & 8.01 (0.06) \\
  & MBSGS & 0.90 (0.03) & 0.30 (0.07) & 0.01 (0.00) & 7.95 (0.06) \\
  & mvSuSiE$^1$ & 0.85 (0.02) & 0.30 (0.07) & 0.01 (0.00) & 7.99 (0.06) \\
  & mvSuSiE$^2$ & 0.83 (0.02) & 0.43 (0.05) & 0.01 (0.00) & *** \\
  & CAVIAR & 0.77 (0.01) & 0.75 (0.04) & 0.01 (0.00) & - \\
  & PAINTOR & 0.52 (0.00) & 0.19 (0.04) & 0.01 (0.00) & - \\
  & remMAP & *** & - & - & *** \\
  & MSGLasso & 0.52 (0.00) & - & - & 7.57 (0.05) \\
 \hline
\end{tabular}
\caption{Simulation results for all methods in Simulation (I)-(V). FDR and FOR are assessed at nominal FDR level of 10\%. Regularization methods including remMAP and MSGLasso do not directly have inference, thus no FDR or FOR are reported. No MSPE values can be directly computed for the fine mapping methods thus we leave them blank. remMAP cannot finish 100 simulations within 2 weeks in high-dimensional case thus we do not report their results for Scenario V. mvSuSiE$^1$: mvSuSiE results based on PIP, mvSuSiE$^1$: mvSuSiE results based on lfsr.}\label{tab:simulations}
\end{table}

\begin{table}[H]
\centering
\small
\begin{tabular}{|c p{0.45cm} c  c  c p{1.4cm}|} 
 \hline
 \textbf{SNP} & \textbf{Chr} &  \textbf{Position} & \textbf{eQTL} & \textbf{Gene} & \textbf{Annotation} \\    
 \hline\hline
rs3829787 & 15 & 78563924 &  No & \textit{CHRNA5} & intergenic \\
rs1824520863 & 15 & 78565470 & No & \textit{CHRNA5} &  - \\
rs503464 & 15 & 78565554 & No & \textit{CHRNA5} & 5'UTR \\
rs55853698 & 15 & 78565597 & No &  \textit{CHRNA5} & 5'UTR \\
rs752882796 & 15 & 78567263 & No &  \textit{CHRNA5} &  intronic \\
rs62010331 & 15 & 78615354 & No & \textit{CHRNA3} & intronic \\
rs3743074 & 15 & 78617138 &  No &  \textit{CHRNA3} &  intronic \\
rs28669908 & 15 & 78617925 &  No &  \textit{CHRNA3} &  intronic \\
rs8040868 & 15 & 78618839 &  No & \textit{CHRNA3} &  exonic \\
rs4366683 & 15 & 78619861 &  No &  \textit{CHRNA3} &  intronic \\
rs149959208 & 15 & 78620368 &  Yes &  \textit{CHRNA3} &  intronic \\
rs1343763852 & 15 &  78621646 &  No &  \textit{CHRNA3} & - \\
 \hline
\end{tabular}
\caption{The causal variants detected by our method and their annotations}\label{tab:real}
\end{table}

\section{Author contributions statement}

T.C. and T.M. conceived the study and developed the method. T.C. wrote the code and the software and conducted the analysis for both simulation and real data application. H.K, S.C. and Z.Y. collected and processed the data in the real data application. T.C. and T.M. wrote the manuscript. All authors reviewed and editted the manuscript.

\section{Acknowledgments}
Research reported in this publication was supported by the National Institute on Drug Abuse (NIDA) of National Institute of Health under the award number 1DP1DA048968-01 to S.C. and T.M., by the University of Maryland MPower Brain Health and Human Performance seed grant and Grand Challenge Grant to T.M.

\markboth{\hfill{\footnotesize\rm TRAVIS CANIDA AND HONGJIE KE AND AND SHUO CHEN AND ZHENYAO YE AND TIANZHOU MA} \hfill}
{\hfill {\footnotesize\rm Travis Canida and Hongjie Ke and SHuo Chen and Zhenyao Ye and Tianzhou Ma} \hfill}

\bibhang=1.7pc
\bibsep=2pt
\fontsize{9}{14pt plus.8pt minus .6pt}\selectfont
\renewcommand\bibname{\large \bf References}
\expandafter\ifx\csname
natexlab\endcsname\relax\def\natexlab#1{#1}\fi
\expandafter\ifx\csname url\endcsname\relax
  \def\url#1{\texttt{#1}}\fi
\expandafter\ifx\csname urlprefix\endcsname\relax\def\urlprefix{URL}\fi


\bibliographystyle{apalike}
\bibliography{Rdc}

\begin{thebibliography}{}

\bibitem[Albert and Chib, 1993]{albert1993bayesian}
Albert, J.~H. and Chib, S. (1993).
\newblock Bayesian analysis of binary and polychotomous response data.
\newblock {\em Journal of the American Statistical Association}, 88(422):669--679.

\bibitem[Bierut et~al., 2008]{bierut2008variants}
Bierut, L.~J., Stitzel, J.~A., Wang, J.~C., Hinrichs, A.~L., Grucza, R.~A., Xuei, X., Saccone, N.~L., Saccone, S.~F., Bertelsen, S., Fox, L., et~al. (2008).
\newblock Variants in nicotinic receptors and risk for nicotine dependence.
\newblock {\em American Journal of Psychiatry}, 165(9):1163--1171.

\bibitem[Biswas et~al., 2022]{biswas2022scalable}
Biswas, N., Mackey, L., and Meng, X.-L. (2022).
\newblock Scalable spike-and-slab.
\newblock In {\em International Conference on Machine Learning}, pages 2021--2040. PMLR.

\bibitem[Borenstein et~al., 2010]{borenstein2010basic}
Borenstein, M., Hedges, L.~V., Higgins, J.~P., and Rothstein, H.~R. (2010).
\newblock A basic introduction to fixed-effect and random-effects models for meta-analysis.
\newblock {\em Research Synthesis Methods}, 1(2):97--111.

\bibitem[Cano-Gamez and Trynka, 2020]{cano2020gwas}
Cano-Gamez, E. and Trynka, G. (2020).
\newblock From gwas to function: using functional genomics to identify the mechanisms underlying complex diseases.
\newblock {\em Frontiers in Genetics}, 11:424.

\bibitem[Casella et~al., 2010]{casella2010penalized}
Casella, G., Ghosh, M., Gill, J., and Kyung, M. (2010).
\newblock Penalized regression, standard errors, and bayesian lassos.
\newblock {\em Bayesian Analysis}, 5(2):369--411.

\bibitem[Cathie et~al., 2015]{sudlow2015ukbiobank}
Cathie, S., John, G., Naomi, A., Valerie, B., Paul, B., John, D., Paul, D., Paul, E., Jane, G., Martin, L., Bette, L., Paul, M., Giok, O., Jill, P., Alan, S., Alan, Y., Tim, S., Tim, P., and Rory, C. (2015).
\newblock Uk biobank: An open access resource for identifying the causes of a wide range of complex diseases of middle and old age.
\newblock {\em PLOS Medicine}, 12(3).

\bibitem[Chen et~al., 2012]{chen2012nicotine}
Chen, L.-S., Xian, H., Grucza, R.~A., Saccone, N.~L., Wang, J.~C., Johnson, E.~O., Breslau, N., Hatsukami, D., and Bierut, L.~J. (2012).
\newblock Nicotine dependence and comorbid psychiatric disorders: examination of specific genetic variants in the chrna5-a3-b4 nicotinic receptor genes.
\newblock {\em Drug and Alcohol Dependence}, 123:S42--S51.

\bibitem[Chen et~al., 2016]{chen2016bayesian}
Chen, R.-B., Chu, C.-H., Yuan, S., and Wu, Y.~N. (2016).
\newblock Bayesian sparse group selection.
\newblock {\em Journal of Computational and Graphical Statistics}, 25(3):665--683.

\bibitem[Consortium et~al., 2015]{gtex2015genotype}
Consortium, G., Ardlie, K.~G., Deluca, D.~S., Segr{\`e}, A.~V., Sullivan, T.~J., Young, T.~R., Gelfand, E.~T., Trowbridge, C.~A., Maller, J.~B., Tukiainen, T., et~al. (2015).
\newblock The genotype-tissue expression (gtex) pilot analysis: multitissue gene regulation in humans.
\newblock {\em Science}, 348(6235):648--660.

\bibitem[Corradin et~al., 2014]{corradin2014combinatorial}
Corradin, O., Saiakhova, A., Akhtar-Zaidi, B., Myeroff, L., Willis, J., Cowper-Sal, R., Lupien, M., Markowitz, S., Scacheri, P.~C., et~al. (2014).
\newblock Combinatorial effects of multiple enhancer variants in linkage disequilibrium dictate levels of gene expression to confer susceptibility to common traits.
\newblock {\em Genome Research}, 24(1):1--13.

\bibitem[Deshpande et~al., 2019]{deshpande2019simultaneous}
Deshpande, S.~K., Ro{\v{c}}kov{\'a}, V., and George, E.~I. (2019).
\newblock Simultaneous variable and covariance selection with the multivariate spike-and-slab lasso.
\newblock {\em Journal of Computational and Graphical Statistics}, 28(4):921--931.

\bibitem[Erzurumluoglu et~al., 2020]{erzurumluoglu2020meta}
Erzurumluoglu, A.~M., Liu, M., Jackson, V.~E., Barnes, D.~R., Datta, G., Melbourne, C.~A., Young, R., Batini, C., Surendran, P., Jiang, T., et~al. (2020).
\newblock Meta-analysis of up to 622,409 individuals identifies 40 novel smoking behaviour associated genetic loci.
\newblock {\em Molecular Psychiatry}, 25(10):2392--2409.

\bibitem[Fang et~al., 2018]{fang2018bayesian}
Fang, Z., Ma, T., Tang, G., Zhu, L., Yan, Q., Wang, T., Celed{\'o}n, J.~C., Chen, W., and Tseng, G.~C. (2018).
\newblock Bayesian integrative model for multi-omics data with missingness.
\newblock {\em Bioinformatics}, 34(22):3801--3808.

\bibitem[Farhad et~al., 2014]{hormozdiari2014caviar}
Farhad, H., Emrah, K., Yong, K.~E., Bogdan, P., and Eleazar, E. (2014).
\newblock Identifying causal variants at loci with multiple signals of association.
\newblock {\em Genetics}, 198(2):497--508.

\bibitem[George and McCulloch, 1993]{george1993variable}
George, E.~I. and McCulloch, R.~E. (1993).
\newblock Variable selection via gibbs sampling.
\newblock {\em Journal of the American Statistical Association}, 88(423):881--889.

\bibitem[Han and Wei, 2013]{han2012Latent}
Han, F. and Wei, P. (2013).
\newblock A composite likelihood approach to latent multivaraite gaussian modeling of snp data with application to genetic association testing.
\newblock {\em Biometrics}, 68(1):307--315.

\bibitem[Harrow et~al., 2012]{harrow2012gencode}
Harrow, J., Frankish, A., Gonzalez, J.~M., Tapanari, E., Diekhans, M., Kokocinski, F., Aken, B.~L., Barrell, D., Zadissa, A., Searle, S., et~al. (2012).
\newblock Gencode: the reference human genome annotation for the encode project.
\newblock {\em Genome Research}, 22(9):1760--1774.

\bibitem[Hern{\'a}ndez-Lobato et~al., 2013]{hernandez2013generalized}
Hern{\'a}ndez-Lobato, D., Hern{\'a}ndez-Lobato, J.~M., Dupont, P., et~al. (2013).
\newblock Generalized spike-and-slab priors for bayesian group feature selection using expectation propagation.
\newblock {\em Journal of Machine Learning Research}.

\bibitem[Hormozdiari et~al., 2016]{hormozdiari2016colocalization}
Hormozdiari, F., Van De~Bunt, M., Segre, A.~V., Li, X., Joo, J. W.~J., Bilow, M., Sul, J.~H., Sankararaman, S., Pasaniuc, B., and Eskin, E. (2016).
\newblock Colocalization of gwas and eqtl signals detects target genes.
\newblock {\em The American Journal of Human Genetics}, 99(6):1245--1260.

\bibitem[Ishwaran and Rao, 2005]{ishwaran2005spike}
Ishwaran, H. and Rao, J.~S. (2005).
\newblock Spike and slab variable selection: frequentist and bayesian strategies.
\newblock {\em The Annals of Statistics}, 33(2):730--773.

\bibitem[Johnstone and Silverman, 2004]{johnstone2004needles}
Johnstone, I.~M. and Silverman, B.~W. (2004).
\newblock Needles and straw in haystacks: Empirical bayes estimates of possibly sparse sequences.
\newblock {\em The Annals of Statistics}, 32(4):1594--1649.

\bibitem[Ke et~al., 2022]{ke2022high}
Ke, H., Ren, Z., Qi, J., Chen, S., Tseng, G.~C., Ye, Z., and Ma, T. (2022).
\newblock High-dimension to high-dimension screening for detecting genome-wide epigenetic and noncoding rna regulators of gene expression.
\newblock {\em Bioinformatics}, 38(17):4078--4087.

\bibitem[Kichaev et~al., 2017]{kichaev2017improved}
Kichaev, G., Roytman, M., Johnson, R., Eskin, E., Lindstroem, S., Kraft, P., and Pasaniuc, B. (2017).
\newblock Improved methods for multi-trait fine mapping of pleiotropic risk loci.
\newblock {\em Bioinformatics}, 33(2):248--255.

\bibitem[Kichaev et~al., 2014]{kichaev2014integrating}
Kichaev, G., Yang, W.-Y., Lindstrom, S., Hormozdiari, F., Eskin, E., Price, A.~L., Kraft, P., and Pasaniuc, B. (2014).
\newblock Integrating functional data to prioritize causal variants in statistical fine-mapping studies.
\newblock {\em PLoS Genetics}, 10(10):e1004722.

\bibitem[Kruschke, 2014]{kruschke2014doing}
Kruschke, J. (2014).
\newblock Doing bayesian data analysis: A tutorial with r, jags, and stan.

\bibitem[Kuo and Mallick, 1998]{kuo1998variable}
Kuo, L. and Mallick, B. (1998).
\newblock Variable selection for regression models.
\newblock {\em Sankhy{\=a}: The Indian Journal of Statistics, Series B}, pages 65--81.

\bibitem[Li and Tseng, 2011]{li2011adaptively}
Li, J. and Tseng, G.~C. (2011).
\newblock An adaptively weighted statistic for detecting differential gene expression when combining multiple transcriptomic studies.
\newblock {\em Annals of Applied Statistics}.

\bibitem[Li et~al., 2021]{li2021joint}
Li, Y., Datta, J., Craig, B.~A., and Bhadra, A. (2021).
\newblock Joint mean--covariance estimation via the horseshoe.
\newblock {\em Journal of Multivariate Analysis}, 183:104716.

\bibitem[Li et~al., 2015]{li2015multivariate}
Li, Y., Nan, B., and Zhu, J. (2015).
\newblock Multivariate sparse group lasso for the multivariate multiple linear regression with an arbitrary group structure.
\newblock {\em Biometrics}, 71(2):354--363.

\bibitem[Liquet et~al., 2017]{liquet2017bayesian}
Liquet, B., Mengersen, K., Pettitt, A., and Sutton, M. (2017).
\newblock Bayesian variable selection regression of multivariate responses for group data.
\newblock {\em Bayesian Analysis}, 12(4):1039--1067.

\bibitem[Liu et~al., 2019]{liu2019association}
Liu, M., Jiang, Y., Wedow, R., Li, Y., Brazel, D.~M., Chen, F., Datta, G., Davila-Velderrain, J., McGuire, D., Tian, C., et~al. (2019).
\newblock Association studies of up to 1.2 million individuals yield new insights into the genetic etiology of tobacco and alcohol use.
\newblock {\em Nature Genetics}, 51(2):237--244.

\bibitem[Ma et~al., 2020]{ma2020variable}
Ma, T., Ren, Z., and Tseng, G.~C. (2020).
\newblock Variable screening with multiple studies.
\newblock {\em Statistica Sinica}, 30(2):925--953.

\bibitem[Mitchell and Beauchamp, 1988]{mitchell1988bayesian}
Mitchell, T.~J. and Beauchamp, J.~J. (1988).
\newblock Bayesian variable selection in linear regression.
\newblock {\em Journal of the American Statistical Association}, 83(404):1023--1032.

\bibitem[Newton et~al., 2004]{newton2004bfdr}
Newton, M., Noueiry, A., Sarkar, D., and Ahlquist, P. (2004).
\newblock Detecting differential gene expression with a semiparametric hierarchical mixture method.
\newblock {\em Biostatistics}, 5(2):155--176.

\bibitem[Nicolae et~al., 2010]{nicolae2010trait}
Nicolae, D.~L., Gamazon, E., Zhang, W., Duan, S., Dolan, M.~E., and Cox, N.~J. (2010).
\newblock Trait-associated snps are more likely to be eqtls: annotation to enhance discovery from gwas.
\newblock {\em PLoS Genetics}, 6(4):e1000888.

\bibitem[O'Hara and Sillanp{\"a}{\"a}, 2009]{o2009review}
O'Hara, R.~B. and Sillanp{\"a}{\"a}, M.~J. (2009).
\newblock A review of bayesian variable selection methods: what, how and which.
\newblock {\em Bayesian analysis}, 4(1):85--117.

\bibitem[Paaby and Rockman, 2013]{paaby2013many}
Paaby, A.~B. and Rockman, M.~V. (2013).
\newblock The many faces of pleiotropy.
\newblock {\em Trends in Genetics}, 29(2):66--73.

\bibitem[Park and Casella, 2008]{park2008bayesian}
Park, T. and Casella, G. (2008).
\newblock The bayesian lasso.
\newblock {\em Journal of the American Statistical Association}, 103(482):681--686.

\bibitem[Peng et~al., 2010]{peng2010regularized}
Peng, J., Zhu, J., Bergamaschi, A., Han, W., Noh, D.-Y., Pollack, J.~R., and Wang, P. (2010).
\newblock Regularized multivariate regression for identifying master predictors with application to integrative genomics study of breast cancer.
\newblock {\em The Annals of Applied Statistics}, 4(1):53.

\bibitem[Quach et~al., 2020]{quach2020expanding}
Quach, B.~C., Bray, M.~J., Gaddis, N.~C., Liu, M., Palviainen, T., Minica, C.~C., Zellers, S., Sherva, R., Aliev, F., Nothnagel, M., et~al. (2020).
\newblock Expanding the genetic architecture of nicotine dependence and its shared genetics with multiple traits.
\newblock {\em Nature Communications}, 11(1):5562.

\bibitem[Saunders et~al., 2022]{saunders2022genetic}
Saunders, G.~R., Wang, X., Chen, F., Jang, S.-K., Liu, M., Wang, C., Gao, S., Jiang, Y., Khunsriraksakul, C., Otto, J.~M., et~al. (2022).
\newblock Genetic diversity fuels gene discovery for tobacco and alcohol use.
\newblock {\em Nature}, 612(7941):720--724.

\bibitem[Schaid et~al., 2018]{schaid2018genome}
Schaid, D.~J., Chen, W., and Larson, N.~B. (2018).
\newblock From genome-wide associations to candidate causal variants by statistical fine-mapping.
\newblock {\em Nature Reviews Genetics}, 19(8):491--504.

\bibitem[Scott et~al., 2016]{scott2016bayes}
Scott, S.~L., Blocker, A.~W., Bonassi, F.~V., Chipman, H.~A., George, E.~I., and McCulloch, R.~E. (2016).
\newblock Bayes and big data: The consensus monte carlo algorithm.
\newblock {\em International Journal of Management Science and Engineering Management}, 11(2):78--88.

\bibitem[Sillanpää and Bhattacharjee, 2005]{sillanpää2005bayesian}
Sillanpää, M.~J. and Bhattacharjee, M. (2005).
\newblock Bayesian association-based fine mapping in small chromosomal segments.
\newblock {\em Genetics}, 169(1):427--439.

\bibitem[Sillanpää and Bhattacharjee, 2006]{sillanpää2006association}
Sillanpää, M.~J. and Bhattacharjee, M. (2006).
\newblock Association mapping of complex trait loci with context-dependent effects and unknown context variable.
\newblock {\em Genetics}, 174(3):1597--1611.

\bibitem[Simon et~al., 2013]{simon2013sparse}
Simon, N., Friedman, J., Hastie, T., and Tibshirani, R. (2013).
\newblock A sparse-group lasso.
\newblock {\em Journal of Computational and Graphical Statistics}, 22(2):231--245.

\bibitem[Stephens, 2013]{stephens2013unified}
Stephens, M. (2013).
\newblock A unified framework for association analysis with multiple related phenotypes.
\newblock {\em PloS one}, 8(7):e65245.

\bibitem[Tibshirani, 1996]{tibshirani1996regression}
Tibshirani, R. (1996).
\newblock Regression shrinkage and selection via the lasso.
\newblock {\em Journal of the Royal Statistical Society: Series B (Statistical Methodology)}, 58(1):267--288.

\bibitem[Wang et~al., 2020]{wang2020simple}
Wang, G., Sarkar, A., Carbonetto, P., and Stephens, M. (2020).
\newblock A simple new approach to variable selection in regression, with application to genetic fine mapping.
\newblock {\em Journal of the Royal Statistical Society: Series B (Statistical Methodology)}, 82(5):1273--1300.

\bibitem[Wang et~al., 2009]{wang2009genetic}
Wang, J., Grucza, R., Cruchaga, C., Hinrichs, A., Bertelsen, S., Budde, J., Fox, L., Goldstein, E., Reyes, O., Saccone, N., et~al. (2009).
\newblock Genetic variation in the chrna5 gene affects mrna levels and is associated with risk for alcohol dependence.
\newblock {\em Molecular Psychiatry}, 14(5):501--510.

\bibitem[Wang et~al., 2007]{wang2007improving}
Wang, T., Zhu, X., and Elston, R.~C. (2007).
\newblock Improving power in contrasting linkage-disequilibrium patterns between cases and controls.
\newblock {\em The American Journal of Human Genetics}, 80(5):911--920.

\bibitem[Wills et~al., 2017]{wills2017phenotypic}
Wills, A.~G., Evans, L.~M., and Hopfer, C. (2017).
\newblock Phenotypic and genetic relationship between bmi and drinking in a sample of uk adults.
\newblock {\em Behavior Genetics}, 47:290--297.

\bibitem[Wills and Hopfer, 2019]{wills2019phenotypic}
Wills, A.~G. and Hopfer, C. (2019).
\newblock Phenotypic and genetic relationship between bmi and cigarette smoking in a sample of uk adults.
\newblock {\em Addictive Behaviors}, 89:98--103.

\bibitem[Xu and Ghosh, 2015]{xu2015bayesian}
Xu, X. and Ghosh, M. (2015).
\newblock Bayesian variable selection and estimation for group lasso.
\newblock {\em Bayesian Analysis}, 10(4):909--936.

\bibitem[Yuan and Lin, 2006]{yuan2006model}
Yuan, M. and Lin, Y. (2006).
\newblock Model selection and estimation in regression with grouped variables.
\newblock {\em Journal of the Royal Statistical Society: Series B (Statistical Methodology)}, 68(1):49--67.

\bibitem[Zhang et~al., 2014]{zhang2014bayesian}
Zhang, L., Baladandayuthapani, V., Mallick, B.~K., Manyam, G.~C., Thompson, P.~A., Bondy, M.~L., and Do, K.-A. (2014).
\newblock Bayesian hierarchical structured variable selection methods with application to molecular inversion probe studies in breast cancer.
\newblock {\em Journal of the Royal Statistical Society: Series C (Applied Statistics)}, 63(4):595--620.

\bibitem[Zhao et~al., 2019]{zhao2019heritability}
Zhao, B., Ibrahim, J.~G., Li, Y., Li, T., Wang, Y., Shan, Y., Zhu, Z., Zhou, F., Zhang, J., Huang, C., et~al. (2019).
\newblock Heritability of regional brain volumes in large-scale neuroimaging and genetic studies.
\newblock {\em Cerebral Cortex}, 29(7):2904--2914.

\bibitem[Zhao et~al., 2021]{zhao2021common}
Zhao, B., Li, T., Yang, Y., Wang, X., Luo, T., Shan, Y., Zhu, Z., Xiong, D., Hauberg, M.~E., Bendl, J., et~al. (2021).
\newblock Common genetic variation influencing human white matter microstructure.
\newblock {\em Science}, 372(6548):eabf3736.

\bibitem[Zhou et~al., 2023]{zhou2023favor}
Zhou, H., Arapoglou, T., Li, X., Li, Z., Zheng, X., Moore, J., Asok, A., Kumar, S., Blue, E.~E., Buyske, S., et~al. (2023).
\newblock Favor: functional annotation of variants online resource and annotator for variation across the human genome.
\newblock {\em Nucleic Acids Research}, 51(D1):D1300--D1311.

\bibitem[Zhu and Zhou, 2020]{zhu2020heritability}
Zhu, H. and Zhou, X. (2020).
\newblock Statistical methods for snp heritability and partition: A review.
\newblock {\em Computational and Structural Biotechnology Journal}, 18:1557--1568.

\bibitem[Zhu et~al., 2019]{zhu2019bayesian}
Zhu, L., Huo, Z., Ma, T., Oesterreich, S., and Tseng, G.~C. (2019).
\newblock Bayesian indicator variable selection to incorporate hierarchical overlapping group structure in multi-omics applications.
\newblock {\em The Annals of Applied Statistics}, 13(4):2611--2636.

\bibitem[Zou and Hastie, 2005]{zou2005regularization}
Zou, H. and Hastie, T. (2005).
\newblock Regularization and variable selection via the elastic net.
\newblock {\em Journal of the Royal Statistical Society: Series B (Statistical Methodology)}, 67(2):301--320.

\bibitem[Zou et~al., 2023]{zou2023mvsusie}
Zou, Y., Carbonetto, P., Xie, D., Wang, G., and Stephens, M. (2023).
\newblock Fast and flexible joint fine-mapping of multiple traits via the sum of single effects model.
\newblock {\em bioRxiv}.

\end{thebibliography}

\end{document}



\markright{ \hbox{\footnotesize\rm Statistica Sinica
}\hfill\\[-13pt]
\hbox{\footnotesize\rm
}\hfill }

\markboth{\hfill{\footnotesize\rm FIRSTNAME1 LASTNAME1 AND FIRSTNAME2 LASTNAME2} \hfill}
{\hfill {\footnotesize\rm Travis Canida and Hongjie Ke and Tianzhou Ma} \hfill}

\renewcommand{\thefootnote}{}
$\ $\par


\fontsize{11}{11pt plus.8pt minus .6pt}\selectfont \vspace{0.8pc}
\begin{center}
\bfseries\large
Multivariate Bayesian variable selection with application to multi-trait genetic fine mapping 
\end{center}

\begin{center}
\normalsize
Travis Canida, Hongjie Ke, Shuo Chen, Zhenyao Ye and Tianzhou Ma
\end{center}

\begin{center}
\textbf{\large Supplementary Material}
\end{center}


\def\thefigure{\arabic{figure}}
\def\thetable{\arabic{table}}

\renewcommand{\theequation}{S\thesection.\arabic{equation}}
\renewcommand{\thefigure}{S\arabic{figure}}
\renewcommand{\thetable}{S\arabic{table}}

\fontsize{11}{11pt plus.8pt minus .6pt}\selectfont

\setcounter{section}{1} 
\setcounter{equation}{0} 
\noindent {\bf 1. Gibbs Sampler}
\label{sec:gibbs}




\par
All the parameters in the model have closed form posterior distributions, thus we use a Gibbs sampler to sample from their posterior distributions. The full posterior distribution of all the unknown parameters conditional on the data can be expressed as:

\begin{equation}
\begin{split}
P& (b,Z,s^2,\Sigma,\alpha,\gamma,\omega, \pi^{(\alpha)},\pi^{(\gamma)},\pi^{(\omega)}  |\textbf{X},\textbf{Y}) \\
& \propto |\Sigma|^{-\frac{n}{2}}\exp{\Biggl\{-\frac{1}{2}Tr\left[\left(\textbf{Y}-\textbf{X} (\textbf{Z}\odot \textbf{b}) \right)\Sigma^{-1}\left(\textbf{Y}-\textbf{X} (\textbf{Z}\odot \textbf{b} )\right)^T\right]\Biggr\}} \\
& \times \prod_{g=1}^G \prod_{j=1}^{m_g} s^{-1}|\Sigma|^{-\frac{1}{2}}\exp\Biggl\{-\frac{1}{2s^2}\vec{b}_{gj}\Sigma^{-1}\vec{b}_{gj}^T\Biggr\} \\
& \times \frac{1}{s^2}  \times |\Sigma|^{-\frac{2q+1}{2}} \exp\Biggl\{-\frac{1}{2}Tr(\textbf{I}\Sigma^{-1})\Biggr\} \\
& \times \prod_{g=1}^G (\pi^{(\alpha)})^{\alpha_g}(1-\pi^{(\alpha)})^{1-\alpha_g} \\
& \times \prod_{g=1}^G \prod_{j=1}^{m_g} (\pi_g^{(\gamma)})^{\gamma_{gj}}(1-\pi_g^{(\gamma)})^{1-\gamma_{gj}} \\
& \times \prod_{g=1}^G \prod_{j=1}^{m_g} \prod_{k=1}^q (\pi_{gj}^{(\omega)})^{\omega_{gj,k}}(1-\pi_{gj}^{(\omega)})^{1-\omega_{gj,k}}
\end{split}
\end{equation}


\noindent\textbf{Conditional posterior distribution}

The conditional posterior distribution of $\vec{b}_{gj}$ is a multivariate normal distribution. Let $\vec{M}_{gj}$ be a $q$-element vector with $k$th element $M_{gj,k}=z_{gj,k}(\textbf{X}_{gj}^T\textbf{X}_{gj}+\frac{1}{s^2})^{-1}\textbf{X}_{gj}^T\textbf{Y}_{(gj),k}$, where $\textbf{X}_{gj}$ is the ($gj$)th column of $\textbf{X}$, $\textbf{Y}_{(gj)}=\textbf{Y}-\textbf{X}_{(gj)}\textbf{B}_{(gj)}$ where $\textbf{X}_{(gj)}$ is the $\textbf{X}$ matrix without ($gj$)th column and $\textbf{B}_{(gj)}$ is the $\textbf{B}$ matrix without ($gj$)th row. Let $\Sigma_{gj}$ be a covariance matrix with diagonal elements $\Sigma_{gj,kk}= \left(\frac{1}{s^2}(\textbf{X}_{gj}^T\textbf{X}_{gj}+\frac{1}{s^2})^{-1}\right)^{z_{gj,k}}s^2\Sigma_{kk}$ and off-diagonal elements $\Sigma_{gj,kk'}=\left(\frac{1}{s^2}(\textbf{X}_{gj}^T\textbf{X}_{gj}+\frac{1}{s^2})^{-1}\right)^{\frac{1}{2}(z_{gj,k}+z_{gj,k'})}s^2\Sigma_{kk'}$ where $k\neq k'$. Then we have that: 
\begin{equation}
    \vec{b}_{gj} | rest \sim N(\vec{M}_{gj},\Sigma_{gj})\\
\end{equation}


The conditional posterior distribution of $s^2$ is an inverse gamma distribution:
\begin{equation}
    s^2|rest \sim IG\left(1+\frac{pq}{2}, \nu+\frac{1}{2}Tr(\textbf{B}\Sigma^{-1}\textbf{B}^T)\right)
\end{equation}

The conditional posterior distributions of the parameters $\pi^{(\alpha)}$, $\pi_g^{(\gamma)}$ and $\pi_{gj}^{(\omega)}$ are all beta distribution: 
\begin{equation}
\begin{split}
\pi^{(\alpha)}|rest & \sim \mathrm{Beta}(1+ \sum\limits_{g=1}^G\alpha_g, 1 + \sum\limits_{g=1}^G (1-\alpha_g)) \\
\pi_g^{(\gamma)}|rest & \sim \mathrm{Beta}(1+ \sum\limits_{j=1}^{m_g} \gamma_{gj}, 1 + \sum\limits_{j=1}^{m_g} (1- \gamma_{gj} ))  \\
\pi_{gj}^{(\omega)}|rest & \sim \mathrm{Beta}(1+ \sum\limits_{k=1}^{q} \omega_{gj,k}, 1 + \sum\limits_{k=1}^{q} (1- \omega_{gj,k} ))  \\
\end{split}
\end{equation}

The conditional posterior distributions of $\alpha_g$ is bernoulli distribution:
\begin{equation}
    \alpha_g|rest \sim \mathrm{Bern}(\pi^{(\alpha)}_*),
\end{equation}
$ \pi^{(\alpha)}_*=\frac{\pi^{(\alpha)}\mathrm{exp}\left\{-\frac{1}{2}\mathrm{Tr}\left[\textbf{X}_g\textbf{B}_{g}\Sigma^{-1}\textbf{B}_{g}^T\textbf{X}^T_g-2\textbf{Y}_{(g)}\Sigma^{-1}\textbf{B}_{g}^T\textbf{X}^T_g\right]\right\}}{\pi^{(\alpha)}\mathrm{exp}\left\{-\frac{1}{2}\mathrm{Tr}\left[\textbf{X}_g\textbf{B}_{g}\Sigma^{-1}\textbf{B}_{g}^T\textbf{X}^T_g-2\textbf{Y}_{(g)}\Sigma^{-1}\textbf{B}_{g}^T\textbf{X}^T_g \right]\right\}+(1-\pi^{(\alpha)})}, $
$\textbf{X}_g$ and $\textbf{B}_{g}$ refer to the $g$th column of $\textbf{X}$ and $g$th row of $\textbf{B}$, respectively. 

The conditional posterior distributions of $\gamma_{gj}$ is also bernoulli: 
\begin{equation}
    \gamma_{gj}|rest \sim \mathrm{Bern}(\pi^{(\gamma)}_{g*}),
\end{equation}
$\pi^{(\gamma)}_{g*}=\frac{\pi^{(\gamma)}_g\mathrm{exp}\left\{-\frac{1}{2}\mathrm{Tr}\left[\textbf{X}_{gj}\textbf{B}_{gj}\Sigma^{-1}\textbf{B}_{gj}^T\textbf{X}^T_{gj}-2\textbf{Y}_{(gj)}\Sigma^{-1}\textbf{B}_{gj}^T\textbf{X}^T_{gj} \right]\right\}}{\pi^{(\gamma)}_g\mathrm{exp}\left\{-\frac{1}{2}\mathrm{Tr}\left[\textbf{X}_{gj}\textbf{B}_{gj}\Sigma^{-1}\textbf{B}_{gj}^T\textbf{X}^T_{gj}-2\textbf{Y}_{(gj)}\Sigma^{-1}\textbf{B}_{gj}^T\textbf{X}^T_{gj}\right]\right\}+(1-\pi^{(\gamma)}_g)}$, $\textbf{X}_{gj}$ and $\textbf{B}_{gj}$ refer to the ($gj$)th column of $\textbf{X}$ and the ($gj$)th row of $\textbf{B}$, respectively.

The conditional posterior distribution of $\omega_{gj,k}$ is also bernoulli: 
\begin{equation}
    \omega_{gj,k}|rest \sim \mathrm{Bern}(\pi^{(\omega)}_{gj*}),
\end{equation}
$ \pi^{(\omega)}_{gj*}=\frac{\pi^{(\omega)}_{gj}\mathrm{exp}\left\{-\frac{1}{2}\mathrm{Tr}\left[\textbf{X}_{gj}\textbf{B}_{gj,k}\Sigma^{-1}_{kk}\textbf{B}_{gj,k}^T\textbf{X}^T_{gj}-2\textbf{Y}_{(gj),k}\Sigma^{-1}_{kk}\textbf{B}_{gj,k}^T\textbf{X}^T_{gj}\right]\right\}}{\pi^{(\omega)}_{gj}\mathrm{exp}\left\{-\frac{1}{2}\mathrm{Tr}\left[\textbf{X}_{gj}\textbf{B}_{gj,k}\Sigma^{-1}_{kk}\textbf{B}_{gj,k}^T\textbf{X}^T_{gj}-2\textbf{Y}_{(gj),k}\Sigma^{-1}_{kk}\textbf{B}_{gj,k}^T\textbf{X}^T_{gj}\right]\right\}+(1-\pi^{(\omega)}_{gj})} $, $\textbf{B}_{gj,k}$ refer to the (gj)th row and kth column of $\textbf{B}$, $\Sigma^{-1}_{kk}$ refers to kth diagonal element of $\Sigma$, $\textbf{Y}_k$ refer to the kth column of $\textbf{Y}$, $\textbf{Y}_{(gj)}=\textbf{Y}-\textbf{X}_{(gj)}\textbf{B}_{(gj)}$ and $\textbf{X}_{(gj)}$ is $\textbf{X}$ without $(gj)$th column and $\textbf{B}_{(gj)}$ is $\textbf{B}$ without $(gj)$th row.

The conditional posterior distribution of $\Sigma$ is inverse-wishart distribution:
\begin{equation}
\Sigma|rest\sim \mathrm{IW}\left(p+q+n,\textbf{I}+(\textbf{Y}-\textbf{X}\textbf{B})^T(\textbf{Y}-\textbf{X}\textbf{B})+s^2\textbf{B}^T\textbf{B}\right)
\end{equation}

To derive the posterior conditional distribution of $d_0$ and $d_1$, we introduce $p$ latent variables $\vec{\xi}=(\xi_{11},\ldots \xi_{Gm_G})^T$ such that $\xi_{gj}|\gamma_{gj},d_0,d_1\sim N(d_0+d_1A_{gj},1)_+$ if $\gamma_{gj}=1$ and $\xi_{gj}|\gamma_{gj},d_0,d_1\sim N(d_0+d_1A_{gj},1)_-$ if $\gamma_{gj}=0$, where $N(\cdot,\cdot)_+$ is the normal distribution truncated at the left by 0, and $N(\cdot,\cdot)_-$ is the normal distribution truncated at the right by 0.

Let $\vec{M}_{d}=(0,\mu_d)^T$ by the prior mean vector and $\Sigma_d$ be the prior covariance matrix with diagonal elements $(\sigma^2,\tau^2)$ and 0 on the off-diagonal. Finally, let $\textbf{A}=\left[\vec{0},\vec{A}\right],\vec{A}=(A_{11},\ldots ,A_{G,m_G})^T$, then, we may update $(d_0,d_1)$ simultaneously from the following distribution:

$$(d_0,d_1)|rest\sim N\left((\Sigma_{d}^{-1}+\textbf{A}^T\textbf{A})^{-1}(\Sigma_{d}\vec{M}_{d}+\textbf{A}^T\vec{\xi}),(\Sigma_{d}^{-1}+\textbf{A}^T\textbf{A})^{-1}\right)$$

To update the $\nu$ in the prior of $s^2$, we use the Monte Carlo EM algorithm as follows: For the $m^{\text{th}}$ EM update, we have

$$\nu^{(m)}=\frac{1}{\mathrm{E}_{\nu^{(m-1)}}\left[\frac{1}{s^2}|\textbf{Y}\right]},$$

where the posterior expectation of $\frac{1}{s^2}$ is estimated from the Gibbs samples based on $\nu^{(m-1)}.$

\section{Derivation of the marginal posterior distribution of $\beta_{gj,k}$}
\label{derivation:marginal}
\begin{proof}
From our model in equation (8) and derivation of Gibbs sampler steps in Section 2.5 of the main text, we have the full conditional posterior distribution for $\vec{b}_{gj}$ given by:
$$\vec{b}_{gj}|rest\sim N(\vec{M}_{gj},\Sigma_{gj}), $$
where $\vec{M}_{gj}$ and $\Sigma_{gj}$ are defined as in equation (14). Under the assumption $\textbf{X}^T\textbf{X}=n\textbf{I}$, we have
$\vec{M}_{gj}=(n+\frac{1}{s^2})^{-1}n\vec{\hat{b}}_{gj}^{LS}$, where $\vec{\hat{b}}_{gj}^{LS}$ is the least squares estimator of $\vec{b}_{gj}$.
 Since $\vec{b}_{gj}$ is multivariate normal, the marginal distribution of $b_{gj,k}$ is also normally distributed:
$$b_{gj,k}|\textbf{X},\textbf{Y}\sim N(M_{gj,k},\Sigma_{gj,kk}).$$

Since $\beta_{gj,k}=z_{gj,k}b_{gj,k}$, we can then obtain the marginal posterior distribution of $\beta_{gj,k}$ in the usual spike-and-slab form as:
$$\beta_{gj,k}|\textbf{X},\textbf{Y}\sim l_{gj,k}N((1-D_{n})\hat{\beta}_{gj,k}^{LS},\frac{1-D_{n}}{n}\Sigma_{kk})+(1-l_{gj,k})\delta_{0}(\beta_{gj,k}),$$
where $D_n=\frac{1}{1+ns^{2}}$ and $l_{gj,k}=\frac{\pi_{gj}^{*} (1+ns^2)^{\frac{1}{2}}\exp{\left[(1-D_n)n\Sigma_{kk}^{-1}(\hat{\beta}_{gj,k}^{LS})^2\right]} }{\pi_{gj}^{*}(1+ns^2)^{\frac{1}{2}}\exp{\left[(1-D_n)n\Sigma_{kk}^{-1}(\hat{\beta}_{gj,k}^{LS})^2\right]} +  (1-\pi_{gj,k}^{*}) }$, and  $\hat{\beta}_{gj,k}^{LS}$ is the least squares estimator of $\beta_{gj,k}$. 
\end{proof}

\section{Derivation of the posterior median as a soft thresholding estimator}
\label{derivation:soft}
\begin{proof}
Define $M=(1-D_n)\hat{\beta}_{gj,k}^{LS}$ and $V=\Sigma_{kk}(1-D_n)/n$. Then, we have the following scenarios:

\begin{itemize}
    \item If $l_{gj,k}<1/2,\hat{\beta}_{gj,k}^{Med}=0$ since $\beta_{gj,k}$ is a mixture of point mass at 0 and $N(M,V)$. 
    \item If $l_{gj,k}\geq 1/2$, then:
    \begin{itemize}
        \item If $\hat{\beta}_{gj,k}^{LS}\geq 0$ and:

        $\Phi(-\frac{\hat{\beta}_{gj,k}^{Med}-M}{\sqrt{V}})=\frac{1}{2l_{gj,k}}\rightarrow \hat{\beta}_{gj,k}^{Med}=(1-D_n)\hat{\beta}_{gj,k}^{LS}-\frac{\sqrt{\Sigma_{kk}}}{\sqrt{n}}\sqrt{(1-D_n)}\Phi^{-1}(\frac{1}{2l_{gj,k}})$

        \item If $\hat{\beta}_{gj,k}^{LS}<0$, then:

        $\Phi(\frac{\hat{\beta}_{gj,k}^{Med}-M}{\sqrt{Var}})=\frac{1}{2l_{gj,k}}\rightarrow \hat{\beta}_{gj,k}^{Med}=(1-D_n)\hat{\beta}_{gj,k}^{LS}+\frac{\sqrt{\Sigma_{kk}}}{\sqrt{n}}\sqrt{(1-D_n)}\Phi^{-1}(\frac{1}{2l_{gj,k}})$
    \end{itemize}
\end{itemize}

If we combine above together, we have:

$\hat{\beta}_{gj,k}^{Med}=sgn(\hat{\beta}_{gj,k}^{LS})((1-D_n)|\hat{\beta}_{gj,k}^{LS}|-\frac{\sqrt{\Sigma_{kk}}}{\sqrt{n}}\sqrt{1-D_n}\Phi^{-1}(\frac{1}{2\mathrm{max}(l_{gj,k},1/2)}))_{+}$,
where $sgn$ is the sign function, $\Phi$ is the cdf of the standard normal distribution, and $(x)_{+}$ takes the value of $x$ if $x>0$, and zero otherwise. 

\end{proof}

\section{Proof of Theorem 2.1}
\label{proof:oracle}
\begin{proof}
Consider $\beta_{gj,k}^0=0$ and $\beta_{gj,k}^0\neq 0$ separately.

\begin{itemize}
    \item When $\beta_{gj,k}^0=0$, from the formula of the soft-thresholding estimator, we have:
$Pr(\hat{\beta}_{gj,k}^{Med}=0|X,Y)=Pr(\frac{\sqrt{1-D_n}\sqrt{n}|\hat{\beta}_{gj,k}^{LS}|}{\sqrt{\Sigma_{kk}}\Phi^{-1}(\frac{1}{2\mathrm{max}(1/2,l_{gj,k}))}}<1)$

    Since $D_n = \frac{1}{1 + ns^2}$, so $D_n \rightarrow 0$; also since $\textbf{X}^TX=n\textbf{I}$ 
 and $\frac{\sqrt{n}|\hat{\beta}_{gj,k}^{LS}-\beta_{gj,k}^0|}{\sqrt{\Sigma_{kk}}}\rightarrow N(0,1)$, thus $\sqrt{n}|\hat{\beta}_{gj,k}^{LS}|=O_p(1)$, and $l_{gj,k}\rightarrow 0$. Thus, we have $Pr(\hat{\beta}_{gj,k}^{Med}=0|\textbf{X},\textbf{Y})\rightarrow 1$. 

    \item When $\beta_{gj,k}^0\neq 0$, we have

    $Pr(\hat{\beta}_{gj,k}^{Med}\neq 0|\textbf{X},\textbf{Y})=Pr(\frac{\sqrt{\Sigma_{kk}}\Phi^{-1}(\frac{1}{2\mathrm{max}(1/2,l_{gj,k})}}{\sqrt{1-D_n}\sqrt{n}|\hat{\beta}_{gj,k}^{LS}|})<1)$.

    By definition, we have 
$$
l_{gj,k}=P(z_{gj,k}=1| rest) = \frac{\pi_{gj}^{*} (1+ns^2)^{\frac{1}{2}}\exp{\left[(1-D_n)n\Sigma_{kk}^{-1}(\hat{\beta}_{gj,k}^{LS})^2\right]} }{\pi_{gj}^{*} (1+ns^2)^{\frac{1}{2}}\exp{\left[(1-D_n)n\Sigma_{kk}^{-1}(\hat{\beta}_{gj,k}^{LS})^2\right]} +(1-\pi_{gj}^{*}) },
$$
where $\pi_{gj}^*=\pi^{(\alpha)} \pi_{g}^{(\gamma)} \pi_{gj}^{(\omega)}$. So we have $1-l_{gj,k} = \frac{ (1-\pi_{gj}^{*}) }{\pi_{gj}^{*} \frac{(1+ns^2)^{1/2}}{\exp(-\frac{1-D_n}{\Sigma_{kk}} n (\hat{\beta}_{gj,k}^{LS})^2 )} +(1-\pi_{gj}^{*}) } $ By L'Hopital's rule, we have: 
$$ \lim_{n\to\infty} \frac{ (1 + ns^2)^{1/2} }{ \exp{\left(-\frac{1-D_n}{2\Sigma_{kk}}n \right)} }  = \lim_{n\to\infty}  \frac{ (s^2/2) (1 + ns^2)^{-1/2} }{ (-\frac{1-D_n}{2\Sigma_{kk}}) \exp{\left(-\frac{1-D_n}{2\Sigma_{kk}}n \right)} }  = 0, $$

since $s^2 / \exp(n) \rightarrow 0 $ and $D_n \rightarrow 0 $. And since $\hat{\beta}_{gj,k}^{LS} \rightarrow \hat{\beta}_{gj,k}^{0} \neq 0 $, by continuous mapping theorem, we have $1- l_{gj,k}\rightarrow 0$ thus $l_{gj,k}\rightarrow 1$. Thus, $Pr(\hat{\beta}_{gj,k}^{Med}\neq 0|\textbf{X},\textbf{Y})\rightarrow 1$




\end{itemize}

\end{proof}

\begin{proof}

We will also prove the asymptotic normality of $\textbf{B}_A^{Med}$.

When $\beta_{gj,k}^0\neq 0$,

\begin{align*}
    & |\sqrt{n}(\hat{\beta}_{gj,k}^{Med}-\hat{\beta}_{gj,k}^{LS})| \\
    &=|\sqrt{n}D_n|\hat{\beta}_{gj,k}^{LS}|+\sqrt{\Sigma_{kk}}\sqrt{1-D_n}\Phi^{-1}(\frac{1}{2\mathrm{max}(1/2,l_{gj,k})})|I(\hat{\beta_{gj,k}^{Med}}\neq 0) \\
    &+\sqrt{n}|\hat{\beta}_{gj,k}^{LS}|I(\hat{\beta}_{gj,k}^{Med}=0)\rightarrow 0
\end{align*}

since $\sqrt{n}D_n\rightarrow 0,\hat{\beta}_{gj,k}^{LS}\rightarrow \beta_{gj,k}^0\neq 0,D_n\rightarrow 0,l_{gj,k}\rightarrow 1,I(\hat{\beta}_{gj,k}^{Med}\neq 0)\rightarrow 1$ and $\sqrt{n}I(\hat{\beta}_{gj,k}^{Med}=0)\rightarrow 0$.

Under orthogonal design, $\sqrt{n}(\hat{\beta}_{A}^{Med}-\hat{\beta}_{A}^{LS})\rightarrow 0$. Since $\sqrt{n}(\hat{\beta}_{A}^{LS}-\beta_{A}^0) \rightarrow N(0,I\otimes \Sigma)$, by Slutsky's theorem, $\sqrt{n}(\hat{\beta}_A^{Med}-\beta_{A}^0)\rightarrow N(0,I\otimes \Sigma)$

\end{proof}

\section{Supplementary Figures and Tables}

\begin{figure}
    \centering
    \includegraphics[scale=0.7]{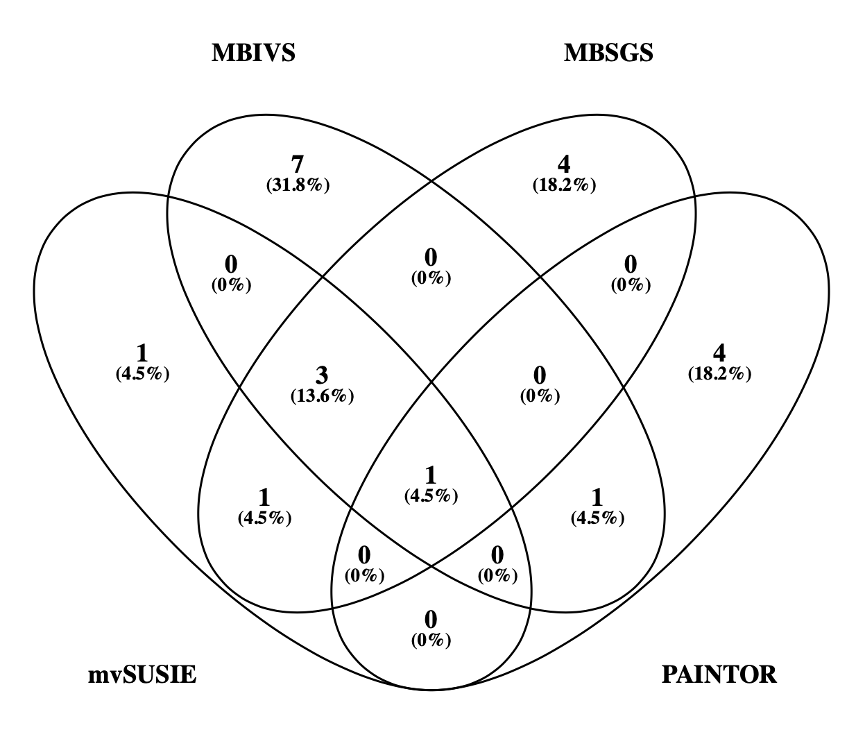}
    \caption{Venn Diagram of detected causal variants by each of the four methods in the real data example.}
    \label{fig:venny}
\end{figure}

\renewcommand{\arraystretch}{0.5}
\begin{table}[ht]
\centering
\small
\begin{tabular}{||c c c c c c||} 
 \hline
 \textbf{Scenario} & \textbf{Model} & \textbf{AUC} & \textbf{FDR} & \textbf{FOR} & \textbf{MSPE} \\    
 \hline\hline
I & mvBaSH & 0.84 (0.01) & 0.65 (0.04) & 0.05 (0.00) & 9.91 (0.05) \\ 
  & MBSGS & 0.67 (0.01) & 0.84 (0.03) & 0.05 (0.00) & 9.92 (0.05) \\
  & mvSuSIE$^1$ & 0.58 (0.01) & 0.48 (0.05) & 0.05 (0.00) & 9.92 (0.05) \\
  & mvSuSIE$^2$ & 0.65 (0.01) & 0.84 (0.03) & 0.05 (0.00) & *** \\
  & CAVIAR & 0.61 (0.01) & 0.87 (0.03) & 0.05 (0.00) & - \\
  & PAINTOR & 0.80 (0.02) & 0.90 (0.03) & 0.05 (0.00) & - \\
  & remMAP & 0.52 (0.00) & - & - & 9.16 (0.00) \\
  & MSGLasso & 0.50 (0.00) & - & - & 9.85 (0.05) \\
II & mvBaSH & 0.89 (0.01) & 0.64 (0.04) & 0.04 (0.00) & 9.94 (0.04) \\
  & MBSGS & 0.80 (0.02) & 0.60 (0.05) & 0.05 (0.00) & 9.94 (0.04) \\
  & mvSuSIE$^1$ & 0.65 (0.02) & 0.45 (0.05) & 0.05 (0.00) & 9.93 (0.04) \\
  & mvSuSIE$^2$ & 0.72 (0.01) & 0.78 (0.04) & 0.05 (0.00) & *** \\
  & CAVIAR & 0.65 (0.01) & 0.88 (0.03) & 0.05 (0.00) & - \\
  & PAINTOR & 0.78 (0.02) & 0.83 (0.04) & 0.05 (0.00) & - \\
  & remMAP & 0.53 (0.00) & - & - & 9.04 (0.05) \\
  & MSGLasso & 0.51 (0.00) & - & - & 9.94 (0.05) \\
III & mvBaSH & 0.84 (0.01) & 0.68 (0.04) & 0.05 (0.00) & 7.90 (0.04) \\
  & MBSGS & 0.65 (0.01) & 0.84 (0.03) & 0.05 (0.00) & 7.90 (0.04) \\
  & mvSuSIE$^1$ & 0.57 (0.01) & 0.47 (0.05) & 0.05 (0.00) & 7.90 (0.04) \\
  & mvSuSIE$^2$ & 0.63 (0.01) & 0.79 (0.04) & 0.05 (0.00) & *** \\
  & CAVIAR & 0.62 (0.01) & 0.93 (0.02) & 0.05 (0.00) & - \\
  & PAINTOR & 0.81 (0.02) & 0.94 (0.02) & 0.05 (0.00) & - \\
  & remMAP & 0.53 (0.01) & - & - & 7.29 (0.04) \\ 
  & MSGLasso & 0.51 (0.00) & - & - & 7.85 (0.04) \\
IV & mvBaSH & 0.88 (0.01) & 0.70 (0.04) & 0.04 (0.00) & 7.92 (0.04) \\
  & MBSGS & 0.79 (0.01) & 0.63 (0.05) & 0.05 (0.00) & 7.92 (0.04) \\
  & mvSuSIE$^1$ & 0.66 (0.02)  & 0.49 (0.05) & 0.05 (0.00) & 7.92 (0.04) \\
  & mvSuSIE$^2$ & 0.70 (0.02) & 0.79 (0.04) & 0.05 (0.00) & *** \\
  & CAVIAR & 0.64 (0.01) & 0.92 (0.02) & 0.05 (0.00) & - \\
  & PAINTOR & 0.80 (0.02) & 0.92 (0.02) & 0.05 (0.00) & - \\
  & remMAP & 0.51 (0.01) & - & - & 7.35 (0.04) \\
  & MSGLasso & 0.50 (0.00) & - & - & 7.85 (0.04) \\
V & mvBaSH & 0.75 (0.03) & 0.93 (0.04) & 0.01 (0.00) & 8.05 (0.08) \\
  & MBSGS & 0.64 (0.02) & 0.95 (0.03) & 0.01 (0.00) & 8.04 (0.08) \\
  & mvSuSIE$^1$ & 0.58 (0.02) & 0.42 (0.08) & 0.01 (0.00) & 8.05 (0.08) \\
  & mvSuSIE$^2$ & 0.70 (0.02) & 0.96 (0.02) & 0.01 (0.00) & *** \\
  & CAVIAR & 0.60 (0.01) & 0.98 (0.01) & 0.01 (0.00) & - \\
  & PAINTOR & 0.77 (0.04) & 0.98 (0.02) & 0.01 (0.00) & - \\
  & remMAP & *** & - & - & *** \\
  & MSGLasso & 0.50 (0.00) & - & - & 7.53 (0.04) \\
 \hline
\end{tabular}
\caption{Simulation results for all methods in scenarios I-V with low heritability (1\%). Regularization methods including remMAP and MSGLasso do not directly have inference, thus no FDR or FOR are reported. No MSPE values can be directly computed for the fine mapping methods thus we leave them blank. remMAP cannot finish 100 simulations within 2 weeks in high-dimensional case thus we do not report their results for Scenario V.}\label{tab:simulations}
\end{table}

\begin{table}[ht]
\centering
\small
\begin{tabular}{||c c c||}
    \hline
    \textbf{Method} & \textbf{Scenario} & \textbf{Computational Cost} \\
    \hline\hline
    mvBaSH & Scenario (I) & 5.51 min \\
     & Scenario (V) & 64.19 min \\
    mvSuSIE & Scenario (I) &  0.33 sec \\
     & Scenario (V) & 0.87 sec \\
    MBSGS & Scenario (I) & 0.94 min \\
     & Scenario (V) & 53.34 min \\
    PAINTOR & Scenario (I) & 3.29 min \\
     & Scenario (V) & $>$ 2 hrs \\
    CAVIAR & Scenario (I) & 10.79 sec \\
     & Scenario (V) & 17.96 min \\
    remMap & Scenario (I) & 21.94 sec \\
     & Scenario (V) & $>$2hrs \\
    MSGLasso & Scenario (I) & 0.42 sec \\
     & Scenario (V) & 10.66 sec \\
    \hline
\end{tabular}
\caption{Average computational cost for running simulation Scenario (I) and (V). For the Gibbs sampling based methods (mvBaSH, MBSGS), the model was fit for 5000 iterations. For penalized methods (remMap, MSGLasso), the time includes finding the best model fit using cross-validation. For mvSuSIE, PAINTOR and CAVIAR, the model was run until convergence was reached. PAINTOR and CAVIAR were used by specifying -enumerate 3 or -c 3 respectively.}
\end{table}


